\documentclass[aps,prd,preprintnumbers,groupedaddress,nofootinbib,amssymb,eqsecnum,notitlepage]{revtex4-1}
\usepackage{here}
\usepackage[dvipdfmx]{graphicx}
\usepackage{amsmath,amsthm,amssymb}
\usepackage{bm}
\usepackage{color}
\usepackage{ulem}

\usepackage{amsfonts}
\usepackage{dcolumn}
\usepackage{hyperref}
\allowdisplaybreaks[1]
\usepackage{stackengine}

\begin{document}

\newcommand{\newc}{\newcommand}
\newcommand{\rk}{\textcolor{red}}
\newc{\be}{\begin{equation}}
\newc{\ee}{\end{equation}}
\newc{\ba}{\begin{eqnarray}}
\newc{\ea}{\end{eqnarray}}
\newc{\D}{\partial}
\newc{\rH}{{\rm H}}
\newc{\rd}{{\rm d}}
\newc{\Mpl}{M_{\rm Pl}}
\newcommand{\rBH}{r_{s}}
\newcommand{\rc}{r_{c}}
\newcommand{\rh}{r_{h}}
\newcommand{\Xh}{X_{h}}
\newcommand{\hr}{\hat{r}}
\newcommand{\ma}[1]{\textcolor{magenta}{#1}}
\newcommand{\cy}[1]{\textcolor{cyan}{#1}}
\newcommand{\mm}[1]{\textcolor{red}{#1}}

\begin{flushright}
WUCG-23-07 YITP-23-69 \\
\end{flushright}

\title{Highly asymmetric probability distribution from a finite-width upward step during inflation}

\author{Ryodai Kawaguchi$^{1,}$\footnote{{\tt ryodai0602@fuji.waseda.jp}}}
\author{Tomohiro Fujita$^{2,3}$}
\author{Misao Sasaki$^{4,5,6}$}

\affiliation{$^{1}$Department of Physics, Waseda University, 3-4-1 Okubo, Shinjuku, Tokyo 169-8555, Japan} 
\affiliation{$^{2}$Waseda Institute for Advanced Study, Waseda University,
1-6-1 Nishi-Waseda, Shinjuku, Tokyo 169-8050, Japan}
\affiliation{$^{3}$Research Center for the Early Universe,  The University of Tokyo, Bunkyo, Tokyo 113-0033, Japan}
\affiliation{$^{4}$Kavli Institute for the Physics and Mathematics of the Universe (WPI),
The University of Tokyo Institutes for Advanced Study,
The University of Tokyo, Chiba 277-8583, Japan}
\affiliation{$^{5}$Center for Gravitational Physics and Quantum Information,
Yukawa Institute for Theoretical Physics, Kyoto University, Kyoto 606-8502, Japan}
\affiliation{$^{6}$Leung Center for Cosmology and Particle Astrophysics,
National Taiwan University, Taipei 10617, Taiwan}

\begin{abstract}
We study a single-field inflation model 
in which the inflaton potential has 
an upward step between two slow-roll regimes by
taking into account the finite width of the step.
We calculate the probability distribution function (PDF) of the curvature perturbation $P[{\cal{R}}]$ using the $\delta N$ formalism. 
The PDF has an exponential-tail only for positive ${\cal{R}}$
whose slope depends on the step width.
We find that the tail may have a significant impact on the estimation of the primordial black hole abundance.
We also show that the PDF $P[{\cal{R}}]$ becomes highly asymmetric on a particular scale exiting the horizon before the step, at which the curvature power spectrum has a dip. This asymmetric PDF may leave an interesting signature in the large scale structure such as voids.
\end{abstract}

\date{\today}


\maketitle


\section{Introduction}
\label{Introduction}
The inflation theory \cite{Starobinsky:1980te,Sato:1980yn,Kazanas:1980tx,Guth:1980zm,Linde:1981mu,Albrecht:1982wi}, originally introduced to resolve the problems of the Big Bang cosmology, is currently accepted as a part of the standard cosmological model.
In particular, slow-roll (SR) inflation driven by a single scalar field produces primordial fluctuations, which are nearly scale-invariant and almost Gaussian, consistent with the observed cosmic microwave background (CMB) temperature anisotropies \cite{Planck:2018vyg,Planck:2018jri}.
Hence, single field SR inflation is considered to be a strong candidate as an inflation model. However, the specific mechanism causing inflation is still unknown, and various inflation models have been proposed within the framework of SR inflation (e.g.\cite{Starobinsky:1980te,Linde:1983gd,Freese:1990rb,Bezrukov:2007ep,Kallosh:2013yoa}).
Even among the canonical single scalar field models, diverse types of inflaton potentials have been studied so far.
Especially at smaller scales than the CMB scale ($k\sim0.002\text{Mpc}^{-1}$), there are only a little observational constraints and many possible inflation scenarios are allowed.

Inflationary models associated with primordial black holes (PBHs) \cite{zel1967hypothesis,Hawking:1971ei,Carr:1974nx,Carr:1975qj} have been intensively studied in recent years (e.g. ultra-slow-roll (USR) models \cite{Garcia-Bellido:2017mdw,Kannike:2017bxn,Germani:2017bcs,Ezquiaga:2017fvi,Motohashi:2017kbs}).
PBHs are black holes formed through the gravitational collapse of regions of high density fluctuations above a threshold in the early universe.
They are hypothetical objects but can play many different roles in the history of the Universe depending on their mass (see \cite{Khlopov:2008qy,Sasaki:2018dmp,Carr:2020xqk,Green:2020jor,Carr:2020gox,Villanueva-Domingo:2021spv,
Carr:2021bzv,Escriva:2022duf,Karam:2022nym,Ozsoy:2023ryl} for recent reviews).
The density fluctuations originate in the curvature perturbation $\cal{R}$ generated during inflation, and the probability of the realisations of large curvature perturbation must be amplified by some mechanism for a sufficient abundance of PBH to be formed.
The simplest way to achieve this is to enhance the power spectrum (i.e. the variance) of the curvature perturbation on a scale small compared to the CMB scale.

In general, however, the power spectrum alone does not completely fix how likely the production of large amplitude curvature perturbations is.
The probabilistic characteristics of the random variable are determined solely by the power spectrum only when it follows a Gaussian distribution.
In discussing deviations from Gaussian distribution in the context of primordial cosmology, higher-order correlators are often used. In particular, the non-gaussian parameter $f_{\rm NL}$ associated with the bispectrum has been discussed extensively~\cite{Komatsu:2001rj,Bartolo:2001cw,Maldacena:2002vr,Acquaviva:2002ud,Bartolo:2004if,Chen:2010xka}.
While SR inflation produces almost Gaussian curvature perturbations characterized by $f_{\rm NL}\ll 1$, the inflation models in which the SR condition is violated can produce relatively large $f_{\rm NL}$ (see e.g. \cite{Namjoo:2012aa,Chen:2013aj,Chen:2013eea,Martin:2012pe,Cai:2018dkf}).
It has been pointed out that the PBH abundance is very sensitive to non-gaussianity in such cases (e.g. \cite{Bullock:1996at,Ivanov:1997ia,Yokoyama:1998pt,Byrnes:2012yx,Bugaev:2013vba,Young:2013oia,Young:2015cyn,Franciolini:2018vbk,Biagetti:2018pjj,Atal:2018neu,Passaglia:2018ixg,DeLuca:2019qsy,Taoso:2021uvl,Davies:2021loj,DeLuca:2022rfz,Matsubara:2022nbr}).

The large and rare fluctuations that cause PBH formation cannot be precisely assessed by perturbative methods alone such as the $f_{\rm NL}$ parameter.
To capture their nature, we can use a powerful non-perturbative approach called the $\delta N$ formalism~\cite{Salopek:1990jq,Sasaki:1995aw,Starobinsky:1985ibc,Sasaki:1998ug,Lyth:2004gb,Lee:2005bb,Lyth:2005fi,Abolhasani:2019cqw}.
This formalism allows a direct conversion from the probability distribution function (PDF) of the scalar field perturbation $\delta\varphi$ to the PDF of $\cal{R}$.
It is known that in inflationary models leading to PBH formation, an exponential tail appears in the PDF of $\cal{R}$, which has a significant impact on the PBH abundance \cite{Atal:2019cdz,Atal:2019erb,Biagetti:2021eep,Kitajima:2021fpq,Gow:2022jfb,Ferrante:2022mui,Pi:2022ysn} (see also stochastic approach \cite{Pattison:2017mbe,Ezquiaga:2019ftu,Figueroa:2020jkf,Pattison:2021oen,Ahmadi:2022lsm,Animali:2022otk}).

This paper focuses on a model where the inflaton potential has an upward step between two SR regions~\cite{Inomata:2021tpx,Cai:2021zsp,Cai:2022erk}.
Models with such an abrupt change have been studied extensively (e.g. \cite{GallegoCadavid:2014jac,Atal:2019cdz,Mishra:2019pzq,Kefala:2020xsx,Inomata:2021uqj,Inomata:2021tpx,Dalianis:2021iig,ZhengRuiFeng:2021zoz,Cai:2021zsp,Cai:2022erk,Kawaguchi:2022nku,Gu:2022pbo,Fu:2022ypp,Cai:2023uhc}) and tend to be consistent with the CMB observations, as the transition takes less time and does not affect much the SR inflaton dynamics on the CMB scale.
Previous studies of the upward step model calculated the PDF of $\cal{R}$ 
by ignoring the finite width of the step and reported that the PDF has a hard cutoff at a certain value of $\cal{R}$
beyond which the PDF is zero~\cite{Cai:2021zsp,Cai:2022erk}.
In this paper, we take a step width $\Delta\varphi$ into consideration.
We introduce a smooth step whose first-order derivative is continuous, and investigate the probabilistic features of the curvature perturbation using the classical $\delta N$ formalism.
We also discuss the impact of the finite step width on the PBH abundance\footnote{Recently, there are active debates if 
the enhancement of the curvature power spectrum on the small scale for PBH
generation could significantly affect the power spectrum on the CMB scale due to the one-loop correction, since Ref.~\cite{Kristiano:2022maq} raised this issue (see also \cite{Inomata:2022yte}).
There are some counter arguments and this is still unsettled issue ~\cite{Choudhury:2023vuj,Choudhury:2023jlt,Choudhury:2023rks,Kristiano:2023scm,Firouzjahi:2023aum,Riotto:2023hoz,Riotto:2023gpm,Firouzjahi:2023ahg,Franciolini:2023lgy,Tasinato:2023ukp,Motohashi:2023syh,Choudhury:2023hvf}. In particular, it is advocated that the loop correction can be sufficiently suppressed by considering smooth transition between SR and USR~\cite{Riotto:2023hoz,Riotto:2023gpm,Firouzjahi:2023ahg}.
We do not address this issue in the present paper, but it gives another motivation for introducing the step width $\Delta\varphi$ that smooths the transition.}. 

We obtain two main results. The first reveals that a tail dependent on the step width $\Delta\varphi$ appears in the PDF, and the cutoff feature disappears.
In the zero-width limit $\Delta\varphi\rightarrow 0$, the tail becomes steeper and eventually the cutoff is reproduced.
The complementary cumulative distribution function is then calculated, showing that the step width $\Delta\varphi$ has a significant impact on the estimate of PBH abundance.
The second is the discovery of a PDF with very high asymmetry.
At certain scales exiting the Hubble horizon before the step, $\cal{R}$ cannot take large negative values and the PDF becomes asymmetric.
The scale that is most asymmetric is shown to correspond to the dip scale, where $\langle ({\cal{R}}-\langle {\cal{R}} \rangle)^{2} \rangle$ and $\langle ({\cal{R}}-\langle {\cal{R}} \rangle)^{3} \rangle$ are the smallest compared to the other scales.
By illustrating the spatial distribution realized with the highly asymmetric PDF, we show that the fraction of low-density regions is smaller at the dip scale.

This paper is organized as follows.
In Sec.~\ref{model}, we explain the details of our model in which the inflaton potential includes an upward step between two SR regions.
In Sec.~\ref{Background}, the background equation of motion is solved, and in Sec.~\ref{deltaN} we derive the relation between the curvature perturbation $\cal{R}$ and the scalar perturbation $\delta\varphi$ using the $\delta N$ formalism.
In Sec.~\ref{PDF}, we compute the PDF and the CCDF of $\cal{R}$ and show that the exponential tail, which depends on $\Delta\varphi$, appears.
In Sec.~\ref{Rmineff} we show that the highly asymmetric PDF appears and relate it to the dip scale. 
We then briefly discuss the implications of the highly asymmetric PDF for the structure of the universe.
Sec.~\ref{conclusion} is devoted to conclusions.

\section{Upward-step model}
\label{model}
In this section, we describe the setup of our model.
Let us consider a single-field inflationary model with a scalar field $\phi$,
\be
\mathcal{S}=\int {\rm d}^{4}x \sqrt{-g} \left[\frac{M_{pl}^{2}}{2}R-\frac{1}{2}g^{\mu\nu}\partial_{\mu}\phi\partial_{\nu}\phi-V(\phi)\right]\,,
\label{action}
\ee
where $g$ is a determinant of the metric tensor $g_{\mu\nu}$ and $V(\phi)$ is a potential of the scalar field.
We consider a spatially flat Friedmann-Lema\^{i}tre-Robertson-Walker background, ${\rm d}s^2=-{\rm d}t^2+a^2(t)\delta_{ij}{\rm d}x^i{\rm d}x^j\,,$ where $a(t)$ is a time-dependent scale factor.
From the above action, we obtain two background equations of motion 
\ba
& &
3h^{2}=\frac{1}{2}h^{2}\pi^{2}+v\,,
\label{friedmann} \\
& &
\frac{{\rm d}\pi}{{\rm d}n}+\frac{v}{h^{2}}\pi+\frac{\partial_\varphi v}{h^{2}}=0\,,
\label{fieldeq}
\ea
where we introduced dimensionless quantities, $\varphi\equiv\phi/M_{pl}$, $\pi\equiv {\rm d}\varphi/{\rm d}n$, $v\equiv V(\phi)/V_0$ and $h\equiv M_{pl} H/\sqrt{V_0}$.
$V_0$ is an arbitrary reference point of the potential, $H=\dot{a}/a$ is the Hubble expansion rate and $n$ is the number of e-folds defined by ${\rm d}n=H{\rm d}t$. 
Without loss of generality, we assume that $\varphi$ moves from positive to zero along the potential and $\pi$ is always negative during inflation.

\begin{figure}[t]
\centering
\includegraphics[width=7.9cm]{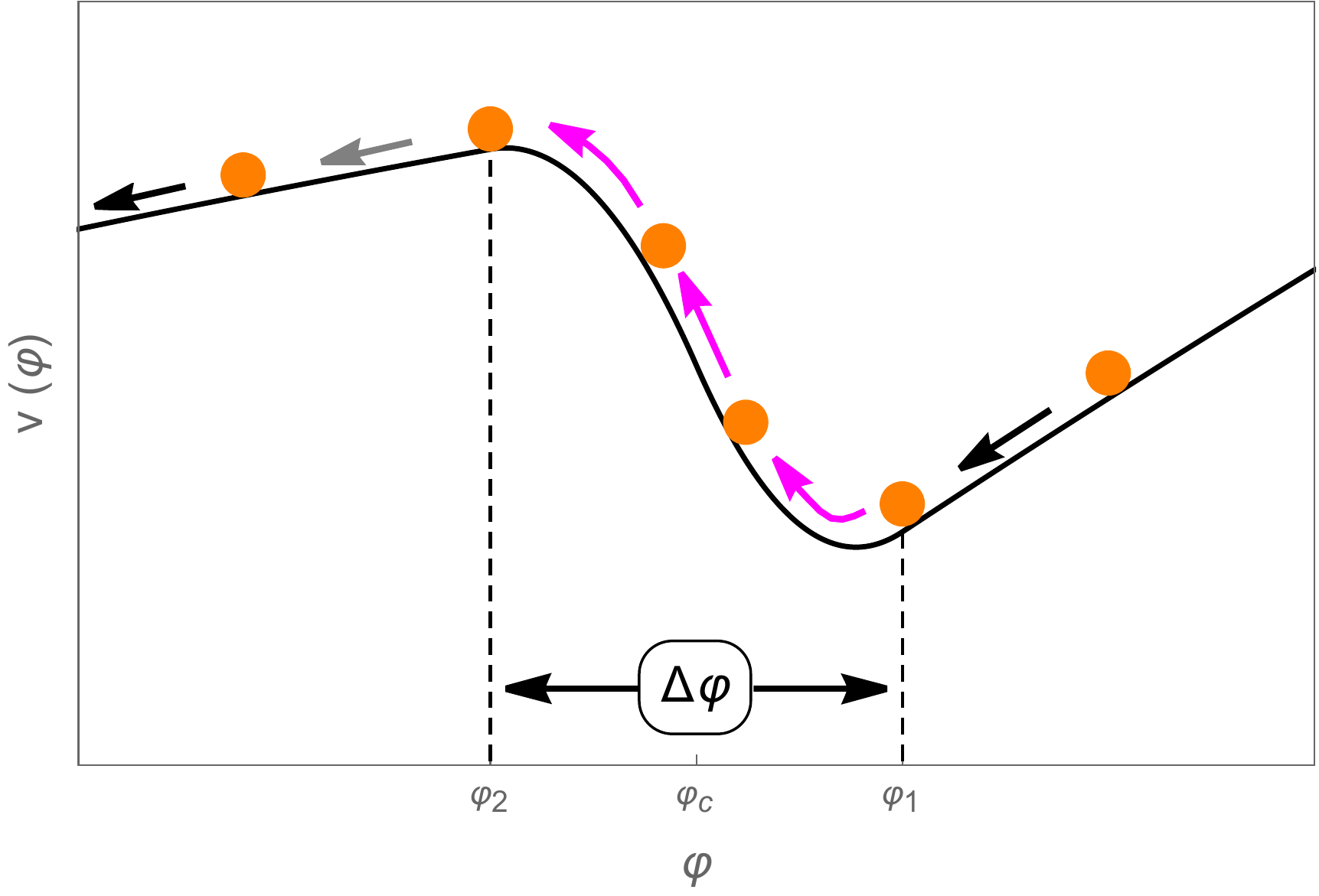}
\hspace{0.8cm}
\includegraphics[width=9cm]{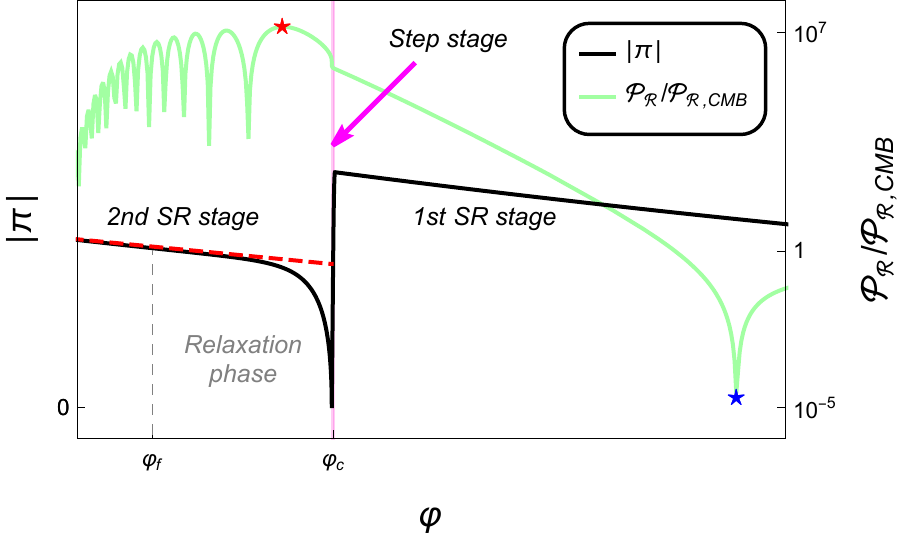}
\caption{\label{stepmodelpot} 
(Left panel) Schematic figure of the potential (black solid line) and the scalar field (orange balls). 
Two slow-roll (SR) potentials are connected by a smooth upward step with the width $\Delta\varphi$.
The black, magenta and gray arrows represent the SR, step-climbing and relaxation phases, respectively.
(Right panel) Schematic figure of the background trajectory in the phase space.
The scalar field evolves from right to left in time, and experiences (i) the first SR stage, (ii) the step stage highlighted by magenta shading, and (iii) the second SR stage. At the beginning of the second SR stage, the trajectory deviates from the SR attractor (red dashed line), which is called the relaxation phase.
For reference, we superimposed the shape of the linear order
power spectrum on the right panel and indicated the positions of its
peak and dip (red and blue stars).
}
\end{figure}

In our model, the potential $v(\varphi)$ has an upward step-like feature between two slow-roll regions,
\be
v(\varphi)=\begin{cases}v_{\rm sr1}(\varphi) \hspace{0.58cm} (\varphi\ge\varphi_{1}) \\ 
f_{\rm step}(\varphi) \hspace{0.44cm} (\varphi_{2}<\varphi<\varphi_{1})\\
 v_{\rm sr2}(\varphi) \hspace{0.575cm} (\varphi\le\varphi_{2})\end{cases}\,,
\label{potential}
\ee
where $v_{\rm sr1}$ and $v_{\rm sr2}$ are slow-roll (SR) potentials, which are generally different from each other and satisfy $v_{\rm sr2}(\varphi_2)>v_{\rm sr1}(\varphi_1)$ to have an upward step in between.
The width of the step region will be an important parameter
\be
\Delta\varphi\equiv\varphi_1-\varphi_2.
\ee
For the upward step $f_{\rm step}$, we employ quadratic functions
\be
f_{\rm step}(\varphi)=\begin{cases}A_1+B_1(\varphi-\varphi_{\rm min})^2\hspace{0.48cm} (\varphi_c\le\varphi<\varphi_{1}: \hspace{0.2cm} \text{S1 region}) \\
A_2+B_2(\varphi-\varphi_{\rm max})^2 \hspace{0.44cm} (\varphi_{2}<\varphi<\varphi_{c} :\hspace{0.2cm} \text{S2 region})\end{cases}\,,
\label{step}
\ee
where $\varphi_c\equiv (\varphi_1+\varphi_2)/2$ is the midpoint of the step region.
$\varphi_{\rm min}$ and $\varphi_{\rm max}$ denote the minimum and maximum of the quadratic potential, respectively.
Requiring that the potential and its first $\varphi$-derivative are continuous at $\varphi=\varphi_c$, $\varphi_1$, and $\varphi_2$, the six constants $A_1,A_2,B_1,B_2,\varphi_{\rm min},\varphi_{\rm max}$ are thereby determined as shown in Appendix.~\ref{paraofstep}.

The left panel of Fig.~\ref{stepmodelpot} schematically shows our potential $v(\varphi)$.
We chose $v_{\rm sr1}$ and $v_{\rm sr2}$ so that their slopes are different before and after the upward step, for example, $\partial_\varphi v_{\rm sr1}(\varphi_{1})> \partial_\varphi v_{\rm sr2}(\varphi_{2})$.
The right panel of Fig.~\ref{stepmodelpot} schematically shows the background trajectory in the phase space.
Due to the presence of the upward step, there are three distinct stages of the background evolution:
(i) The first stage, $\varphi\ge\varphi_1$, is the first SR stage where the background trajectory is on the SR attractor. 
(ii) The second, $\varphi_2<\varphi<\varphi_1$, is the step stage.
In the step stage, the scalar field rapidly loses its kinetic energy and $|\pi|$ becomes smaller than before the step. 
(iii) The third, $\varphi\le\varphi_2$, is the second SR stage.
Since $|\pi|$ is made smaller by the step, the trajectory in the phase space is not on the SR attractor right after the scalar field finishes climbing up the step.
Therefore, after the step, a relaxation phase takes place and the scalar field asymptotically approaches to the second SR attractor.
Eventually, the trajectory can be regarded as settling on the SR attractor, and we call the scalar field value at that time as $\varphi_f$. 

For later convenience, we conclude this section by defining useful quantities as follows.
\ba
&&
\eta_i\equiv2(2\epsilon_{Vi}-\eta_{Vi})\,,\\
&&
g\equiv\frac{\pi_2}{\pi_1}<1\,,\\
&&
\kappa\equiv\sqrt{\frac{\epsilon_{V1}}{\epsilon_{V2}}}\,,\label{kappa}\\
&&
\omega_{\rm s1}\equiv\sqrt{\frac{6B_1}{A_1}}\simeq\frac{\sqrt{2}~ |\pi_1|}{\Delta\varphi}\,,\label{omegas1}\\
&&
\omega_{\rm s2}\equiv\sqrt{\frac{-6B_2}{A_2}}\simeq\frac{\sqrt{2}~ |\pi_1|}{\Delta\varphi}\,,\label{omegas2}
\ea
where  $\epsilon_{V}\equiv (\partial_{\varphi}v/v)^2/2$ and $\eta_{V}\equiv \partial_{\varphi\varphi} v/v$ are the potential slow-roll parameters and the subscript $i$ indicates that a quantity is evaluated at $\varphi=\varphi_i$ ($i=1,2$).
The approximations used in Eqs.~\eqref{omegas1} and \eqref{omegas2} assume 
$\Delta\varphi\ll|\pi_1|$ and $g\ll1$, and the dependence of $\omega_{\rm s1}$
and $\omega_{\rm s2}$ on $\Delta \varphi$ will be emphasized in the later discussion.

\section{Background solution}
\label{Background}
In this section, we briefly discuss the background behavior of the scalar field in each of the three stages. In solving the background equation of motion (\ref{fieldeq}), we use the following three approximations. 
(I) The kinetic energy of the scalar field is subdominant compared to the potential energy, $h^2\pi^2\ll v$.
(I\hspace{-1.2pt}I) Assuming the width of the step $\Delta \varphi$ is sufficiently small, we keep only its leading contributions. Consequently, $\varphi_{\rm min}$ and $\varphi_{\rm max}$ are identified with $\varphi_1$ and $\varphi_2$, respectively.
(I\hspace{-1.2pt}I\hspace{-1.2pt}I) The Hubble friction is negligible in the step stage.
The detailed derivations of the solutions and the discussions on the limitations of the above approximations can be found in Appendices.~\ref{solvingBGEOM} and \ref{Hubblefriction}.

In the SR stages, a general SR solution is applicable,
\be
\varphi(n)-\varphi_i=\frac{2\sqrt{2\epsilon_{Vi}}}{\eta_i}\left(1-e^{\frac{\eta_{i}}{2}(n-n_i)}\right)+\frac{1}{3}\left(\pi_i+\sqrt{2\epsilon_{Vi}}\right)\left(e^{\frac{\eta_{i}}{2}(n-n_i)}-e^{-(3+\frac{\eta_i}{2})(n-n_i)}\right)\,.
\label{BGsolSR}
\ee
Again, we used the shorthand notation $X_i\equiv X(\varphi_i)$.
One needs to know $\pi_i$ to fix the boundary condition.
In the first SR stage, the background trajectory is on the SR attractor, i.e. $\pi_1=-\sqrt{2\epsilon_{V1}}$ and hence the second term in Eq.~\eqref{BGsolSR} vanishes.
In the second SR stage, however, the trajectory is not on the attractor in the relaxation phase (see Fig.~\ref{stepmodelpot}).
To find $\pi_2$ we use the approximated energy conservation during the step stage,
\be
\pi_2=-\sqrt{\pi_1^2+6\log\left(\frac{v(\varphi_1)}{v(\varphi_2)}\right)}\,.
\label{pi2}
\ee
Here $\pi^2/2+3\log(v)$ is conserved, because the background equation of motion with the above approximations (I) and (I\hspace{-1.2pt}I\hspace{-1.2pt}I) reduces to ${\rm d}\pi/{\rm d}n+3\partial_\varphi \log(v)= 0$ as shown in Eq.~\eqref{appfieldeq}.

The step stage is split into the S1 and S2 region, which has a normal and inverted harmonic potential, respectively (see Eq.~\eqref{step}). The background solutions in these regions are 
\ba
&& \varphi(n)-\varphi_{1}=\frac{\pi_{1}}{\omega_{\rm s1}} \sin\left(\omega_{\rm s1}(n-n_1)\right)\,, \qquad (\varphi_c\le\varphi<\varphi_{1})
\label{BGsolstep1} \\
&& \varphi(n)-\varphi_{2}=\frac{\pi_{2}}{\omega_{\rm s2}} \sinh\left(\omega_{\rm s2}(n-n_2)\right)\,.\qquad (\varphi_{2}<\varphi<\varphi_{c})
\label{BGsolstep2}
\ea
Note that for extremely small $\pi_2$, Eq.~\eqref{BGsolstep2} can be inverted
as $\omega_{s2}(n_2-n)\simeq \log(2\omega_{s2}(\varphi(n)-\varphi_2)/|\pi_2|)$.
This implies that it takes logarithmically longer time for the background inflaton to climb up the upward step for smaller $\pi_2$.
It will have a significant impact on the final result.

\section{Calculation of {\boldmath$\delta N$}}
\label{deltaN}
We now turn our attention to the the curvature perturbation $\cal{R}$.
According to the $\delta N$ formalism, the curvature perturbation $\cal{R}$ can be calculated as the difference in the number of e-folds between the perturbed spacetime and the background spacetime \cite{Salopek:1990jq,Sasaki:1995aw,Starobinsky:1985ibc,Sasaki:1998ug,Lyth:2004gb,Lee:2005bb,Lyth:2005fi},
\be
{\cal{R}}=\delta N\equiv
N(\varphi+\delta\varphi,\pi+\delta\pi;\varphi_f,\pi_f)-N(\varphi,\pi;\varphi_f,\pi_f),
\label{deltaNformula}
\ee
where $N(\varphi,\pi;\varphi',\pi')$ is the number of e-folds for which the inflaton takes to evolve from $(\varphi,\pi)$ to $(\varphi',\pi')$ in the phase space. In this paper, we assign $(\varphi,\pi)$ to the background value at the horizon crossing time for a scale of interest $k$.
Note that the velocity perturbation $\delta\pi$ at the starting point is assumed to be negligibly small.
In what follows, we start from $\varphi$ on the first SR attractor ($\varphi>\varphi_1$),
and compute $\delta N$ generated in all stages
\footnote{
The e-folding number taken for the inflaton to pass through the step stage is only about $\Delta N\sim1/\omega_{\rm s1}\ll 1$ and the corresponding band of wavenumbers
that exit the Hubble horizon during the step stage, is very narrow. 
Thus, we do not consider curvature perturbations on such scales in this paper.
The scales crossing the Hubble horizon after the step stage are not studied in this paper either, as they are similar to the normal SR case, 
although the evolution of $\delta\varphi$ well inside the horizon is affected by the step.}.

In the first SR stage, the initial perturbation $\delta\varphi$ induces $\delta N$ as well as the velocity perturbation at the end of the first stage $\delta\pi_1$.
We obtain them by comparing Eq.~\eqref{BGsolSR} at $i=1$ to
its perturbed expression with $\varphi\to \varphi + \delta\varphi$ and $\pi_1 \to \pi_1 + \delta \pi_1$ as (see Appendix.\ref{apppSR1} for derivation)
\be
\delta N^{(1)}\simeq -\frac{\delta\varphi}{\pi} \,,
\qquad
\delta\pi_1\simeq -\frac{\eta_1}{2}\left(\frac{\pi}{\pi_1}\right)^{\frac{6}{\eta_1}}\delta\varphi\,,
\label{deltaN1}
\ee
where $\delta N^{(1)}$ denotes the contribution to total $\delta N$ from the first SR stage.
This approximation is valid for $|\delta N^{(1)}|\ll 1/3$.
One can apparently choose $\varphi$ from any values within the first SR stage.
However, since $6/\eta_1$ is large, if we choose $\varphi$ far from the step, $\delta\pi_1$ would be substantially suppressed by the factor $(\pi/\pi_1)^{6/\eta_1}$.
This is a consequence of the fact that the background trajectory is an attractor in the first SR stage.
If $\delta\pi_1$ is strongly suppressed, there would be no significant contribution to total $\delta N$ from the subsequent evolution in the step stage.
In this paper, therefore, we consider $\varphi$ close to $\varphi_1$ and investigate the effect of the upward step on the curvature perturbations,
that exit the horizon slightly before entering the step stage.

In the second SR stage, $\delta N$ is produced by $\delta\pi_2$, contrary to $\delta N^{(1)}$ induced by $\delta\varphi$.
Perturbing Eq.~(\ref{pi2}), we find that $\delta\pi_2$ is given by
\be
\delta\pi_2=\pi_2\left(\sqrt{1+\frac{2}{g^2}\frac{\delta\pi_1}{\pi_1}+\frac{1}{g^2}\left(\frac{\delta\pi_1}{\pi_1}\right)^2}-1\right)\,,
\label{deltapi2}
\ee
where the Hubble friction in the step stage was ignored.
Using Eq.~\eqref{BGsolSR} at $i=2$ in a similar way to Eq.~\eqref{deltaN1} but only with $\pi_2\to \pi_2+\delta \pi_2$, we obtain the contribution to $\delta N$ from the second SR stage as (see Appendix.\ref{pRelaxation})
\be
\delta N^{(2)}\simeq
-\frac{\kappa g}{3}\frac{\delta\pi_2}{\pi_2}\simeq \frac{\kappa g}{3}\left(1-\sqrt{1+\frac{2}{g^2}\frac{\delta\pi_1}{\pi_1}+\frac{1}{g^2}\left(\frac{\delta\pi_1}{\pi_1}\right)^2}\right)\,.
\label{deltaN22}
\ee
Note that the first approximate equality in Eq.~(\ref{deltaN22}) 
relies only on $|1/\eta_2|\gg 1$ and we do not assume that $\delta\pi_2/\pi_2$ is small.
Indeed, we will consider the case with $|\delta\pi_2/\pi_2|=\mathcal{O}(1)$ soon below.

The contribution to $\delta N$ from the step stage is divided into two: one from the S1 region $\delta N^{(\rm s1)}$ and the other from the S2 region $\delta N^{(\rm s2)}$.
The former is highly suppressed by $\Delta\varphi$ and $(\pi/\pi_1)^{6/\eta_1}$,
and negligible compared to $\delta N^{(1)}$ (see Appendix.\ref{pStep}).
In contrast, $\delta N^{(\rm s2)}$ can be significant. As discussed below Eq.~\eqref{BGsolstep2},
the background e-folds of the S2 region is given by
\be
N^{\rm (s2)}\simeq\frac{1}{\omega_{\rm s2}}\sinh^{-1}\left(\frac{\Delta\varphi}{2|\pi_2|}\omega_{\rm s2}\right)
\simeq\frac{1}{\omega_{\rm s2}}\log\left(\frac{\Delta\varphi}{|\pi_2|}\omega_{\rm s2}\right)\,.
\ee
Perturbing it, we obtain 
\be
\delta N^{\rm (s2)}\simeq-\frac{1}{\omega_{\rm s2}}\log\left(1+\frac{\delta\pi_2}{\pi_2}\right)\,.
\label{deltaNs2}
\ee
It is important to note that when $\delta\pi_2$ is comparable to $-\pi_2$, $\delta N^{\rm (s2)}$ may diverge to infinity. 
This corresponds to the perturbed cases in which the inflaton exhausts the most of the kinetic energy to climb up the step and takes an enormous amount of time to pass through it. 
As we will see below, $\delta N^{\rm (s2)}$ 
makes a significant impact on the curvature perturbation.

Summing up the contributions to $\delta N$ from the three stages, Eqs.~\eqref{deltaN1},\eqref{deltaN22} and \eqref{deltaNs2}, we obtain
\be
{\cal{R}}=\beta \delta\varphi+\frac{\kappa g}{3}\left(1-\sqrt{1+\frac{2\gamma}{g^2}\delta\varphi+\frac{\gamma^2}{g^2}\delta\varphi^2}\right)-\frac{1}{2\omega_{\rm s2}}\log\left(1+\frac{2\gamma}{g^2}\delta\varphi+\frac{\gamma^2}{g^2}\delta\varphi^2\right)\,,
\label{R}
\ee
where $\beta$ and $\gamma$ are given by
\ba
\beta=-\frac{1}{\pi}>0\,,
\qquad
\gamma=\frac{\eta_1\beta}{2}\left(\frac{\pi}{\pi_1}\right)^\frac{6}{\eta_1}\simeq\frac{\eta_1\beta}{2}\left(\frac{k}{k_1}\right)^3\,.
\label{gamma}
\ea
The second and third terms in Eq.~(\ref{R}) originate from the upward step.
The third term particularly comes from the finite width of the step.
In fact, vanishing the step width $\Delta\varphi\to 0$ (i.e. $\omega_{\rm s2}\to \infty$) and dropping the third term would reproduce the results of previous papers that ignored the width of the step~\cite{Cai:2021zsp,Cai:2022erk}.
Instead, substituting $g=1$ into Eq.~(\ref{R}), 
we obtain a similar logarithmic dependence on $\delta \varphi$ to the result in the bump-type model~\cite{Atal:2019cdz}\footnote{
In the limit $g\rightarrow 1$ while keeping $\Delta\varphi$ finite, our step potential becomes
a bump feature, not the featureless SR potential. The previous paper~\cite{Atal:2019cdz} computed $\delta N$ only in the bump region. Hence, it is reasonable that the logarithmic term in Eq.~\eqref{R} approaches to Eq.~(1.5) in the previous paper \cite{Atal:2019cdz} for $g=1$ and $\varphi=\varphi_1$. However, its coefficient is not exactly the same but differs by $\sqrt{3}$, probably because our assumption (I\hspace{-1.2pt}I) i.e. $\varphi_{\rm max}\simeq\varphi_2$ and $\varphi_{\rm min}\simeq\varphi_1$, is broken in the limit $g\to 1$.}

\section{Probability distribution function of {\boldmath$\cal{R}$}}
\label{PDF}
\begin{figure}[t]
\centering
\includegraphics[width=8cm]{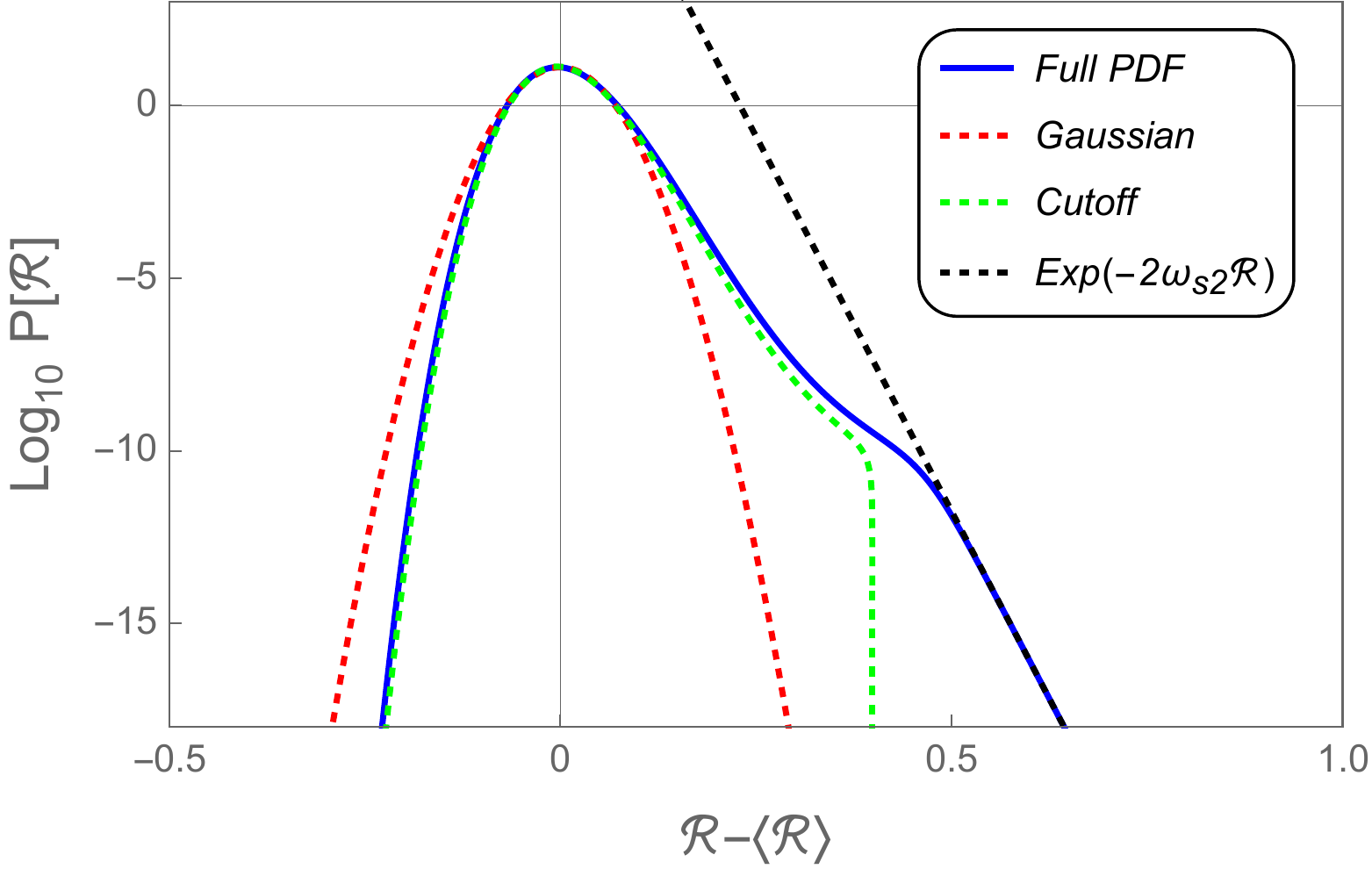}
\hspace{1cm}
\includegraphics[width=8cm]{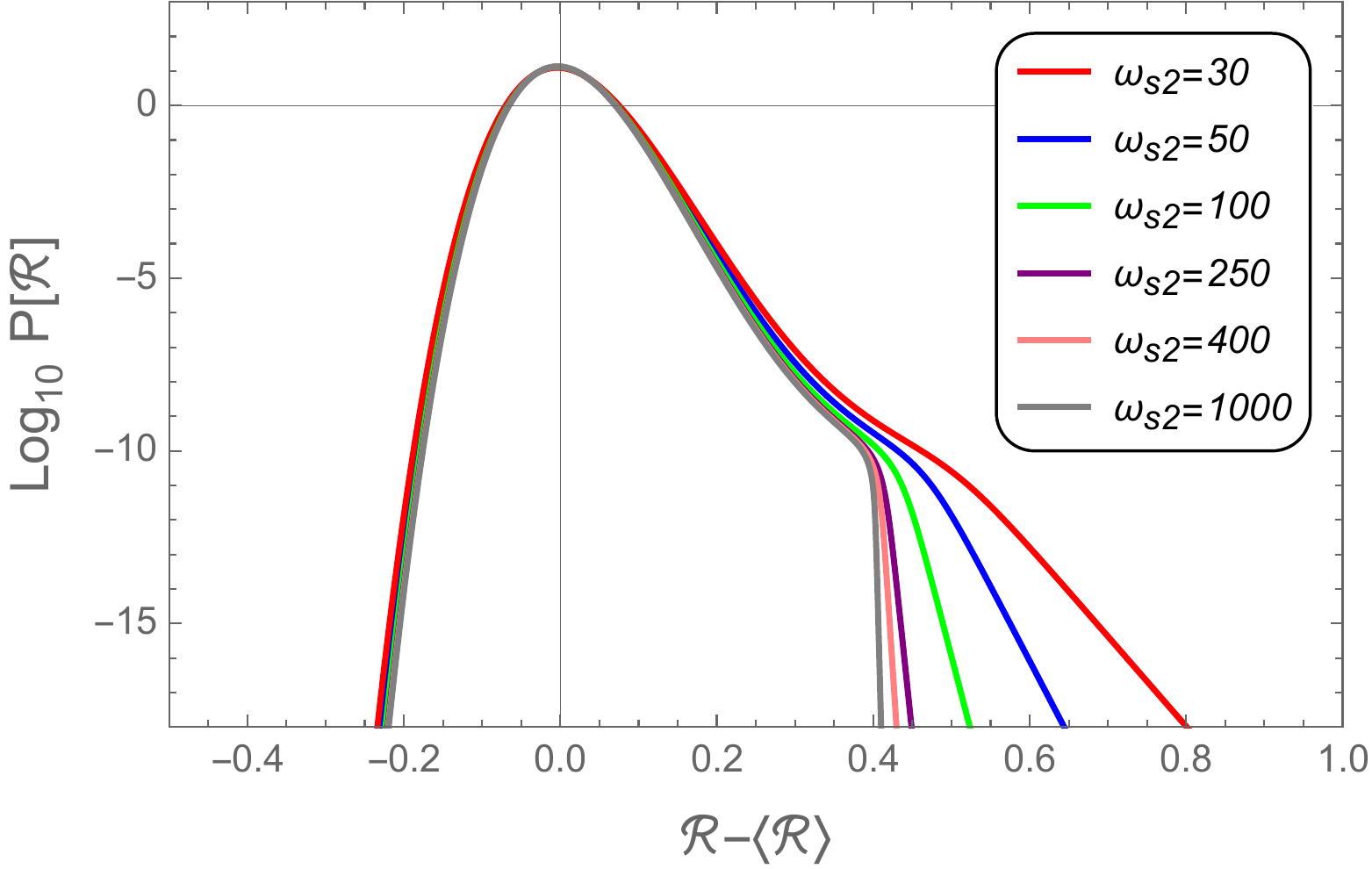}
\caption{The probability distribution function (PDF) of the curvature perturbation $\cal{R}$ against $\cal{R}-\langle \cal{R}\rangle$ for $\gamma=1.5$, $\kappa=60$, $\beta=20$, $g=0.02$, and $\sigma_{\delta\varphi}=2\times10^{-5}$.
The mean value $\langle {\cal{R}}\rangle= \mathcal{O}(10^{-3})$ is negligibly small.
(Left panel) 
The solid blue line denotes the full PDF obtained numerically based on Eq.~\eqref{PDFofR} for $\omega_{\rm s2}=50$. 
The coloured dotted lines represent the Gaussian (red), cutoff (green) and exponential tail (black) given in Eq.~\eqref{eachP}.
(Right panel) The coloured lines denote the full PDFs for $\omega_{\rm s2}=30, 50, 100, 250, 400$ and $1000$ from top to bottom.
They agree with the analytical behavior of the tail $P[{\cal{R}}]\propto \exp(-2\omega_{\rm s2}{\cal{R}})$.
Since $\omega_{\rm s2}\propto 1/\Delta\varphi$, a steeper step tends to produce a steeper tail.
}
\label{PDFex}
\end{figure}

The second and third terms in Eq.~\eqref{R} make the curvature perturbation highly non-Gaussian.
To investigate the properties of $\cal{R}$, we calculate its probability distribution function (PDF).
The PDF of $\cal{R}$ is associated with the PDF of $\delta\varphi$ by probability conservation, $P[{\cal{R}}]=P[\delta\varphi]|{\rm d}\delta\varphi/{\rm d}{\cal{R}}|$.
For simplicity, we assume that $\delta\varphi$ follows a Gaussian distribution with variance $\sigma_{\delta\varphi}^2$.\footnote{We mention that $\delta\varphi$ may have a non-Gaussianity of non-local type, arising from non-trivial higher order interactions. However, since such a non-Gaussianity is known to be perturbatively small in comparison with the local type non-Gaussianity for a canonical scalar field \cite{Chen:2008wn}, we ignore it.
Considering a possible non-perturbative effect of a non-local type non-Gaussianity is an important issue, but it is beyond the scope of the present paper.}
From Eq.~\eqref{R}, the PDF of $\cal{R}$ is given by
\be
P[{\cal{R}}]={\displaystyle \sum_{i}}\frac{\frac{1}{\sqrt{2\pi\sigma_{\delta\varphi}^2}}\left(1+\frac{2\gamma}{g^2}\delta\varphi_i+\frac{\gamma^2}{g^2}\delta\varphi_i^2\right)\exp\left(-\frac{\delta\varphi_i^2}{2\sigma_{\delta\varphi}^2}\right)}{\left|\beta\left(1+\frac{2\gamma}{g^2}\delta\varphi_i+\frac{\gamma^2}{g^2}\delta\varphi_i^2\right)-\frac{\gamma}{g^2}\left(1+\gamma\delta\varphi_i\right)\left(\frac{\kappa g}{3}\sqrt{1+\frac{2\gamma}{g^2}\delta\varphi_i+\frac{\gamma^2}{g^2}\delta\varphi_i^2}+\frac{1}{\omega_{\rm s2}}\right)\right|}\,,
\label{PDFofR}
\ee
where $\delta\varphi_i$'s ($i=a,b$) are the two solutions of Eq.~(\ref{R}) for a given $\cal{R}$\footnote{Strictly speaking there is the possibility that Eq.~(\ref{R}) has another solution for a given $\cal{R}$ in a region $\delta\varphi<-2/\gamma$.
In the following, the contribution of this solution to the probability is ignored as being small.}.
Numerically computing the above equation, we illustrate the PDF of $\cal{R}$ in the left panel of Fig.~\ref{PDFex}.
One can clearly see an asymmetric non-gaussian shape of the full PDF.

To understand several features of the obtained PDF, it is useful to compare it with simplified PDFs, which take into account only one term out of three in Eq.~(\ref{R}).
If one term gives a dominant contribution, an analytic expression of the PDF is available.
Note that for the first term, which is a linear term of $\delta\varphi$, we linearize the other terms and collect all the contributions.
We find that only with the linearized, second, third term in Eq.~\eqref{R}, the PDF would be 
(see Appendix.~\ref{appeachPDF} for derivation)
\be
P[{\cal{R}}]\propto
\begin{cases}
\exp\left[-\frac{{\cal{R}}^2}{2\left(\beta-\frac{\kappa\gamma}{3g}-\frac{\gamma}{\omega_{\rm s2}g^2}\right)^2\sigma_{\delta\varphi}^2}\right]\hspace{4.4cm} :(\text{Gaussian}) \\
\left(1-\frac{3{\cal{R}}}{g\kappa}\right)\exp\left[-\frac{1}{2\sigma_{\delta\varphi}^2}\frac{9g^2{\cal{R}}^2}{\gamma^2\kappa^2}\left(1-\frac{3}{2}\frac{{\cal{R}}}{g\kappa}\right)^2\right] \hspace{2.35cm}  :(\text{Cutoff}) \\
\exp\left(-2\omega_{\rm s2}{\cal{R}}\right)\exp\left[-\frac{1}{2\sigma_{\delta\varphi}^2}\frac{g^2}{4\gamma^2}\Bigl(\exp\left(-2\omega_{\rm s2}{\cal{R}}\right)-1\Bigr)^2\right] \hspace{0.65cm}  :(\text{Exponential tail})
\end{cases}
\label{eachP}
\ee
where $g\ll 1$ is assumed for simplicity.
We dubbed these three cases as Gaussian, Cutoff and Exponential tail after their features.
The linearized term yields a Gaussian PDF. 
The PDF of the second term exhibits a sharp cutoff at ${\cal{R}}={\cal{R}}_{\rm cutoff}\equiv\kappa g/3$ due to its prefactor.
The PDF of the third term characterized by its exponential tail, $\exp(-2\omega_{\rm s2}{\cal{R}})$.

In the left panel of Fig.~\ref{PDFex}, these three simplified PDFs in Eq.~\eqref{eachP} are shown as dotted coloured lines in comparison with the full PDF.
The full PDF (blue solid) matches the Gaussian one (red dotted) only for small $\cal{R}-\langle \cal{R}\rangle$ and their discrepancy quickly gets significant.
Since we chose large $\kappa=60$ in this figure, the second term in Eq.~\eqref{R}
is dominant for a wide region, and the full and
the cutoff PDF (green dotted) are well overlapped up to the cutoff position $\cal{R}-\langle \cal{R}\rangle \simeq {\cal{R}}_{\rm cutoff} =$ 0.4.
For larger $\cal{R}$, the exponential tail from the third term (black dotted) becomes prominent, and the full PDF follows it as expected.
Therefore, the cutoff feature in the PDF advocated in the previous works~\cite{Cai:2021zsp,Cai:2022erk} is covered up by the tail contribution as long as the finite width of the step is correctly taken into account.

To scrutinize the tail behavior, we present several PDFs with varying $\omega_{\rm s2}$ in the right panel of Fig.~\ref{PDFex}.
Recalling that $\omega_{\rm s2}\propto1/\Delta\varphi$ and the analytic tail PDF $P[{\cal{R}}]\propto \exp(-2\omega_{\rm s2}{\cal{R}})$, we expect the slope of the tail becomes steeper as $\Delta\varphi$ decreases (i.e. the step is steeper).
The figure confirms this behavior.
For very large $\omega_{\rm s2}$, the cutoff of the PDF appears to be reproduced.
However, it should be stressed that it is not a hard cutoff but a steep exponential tail. Thus, the probability density is still non-zero at $\cal{R}\ge {\cal{R}}_{\rm cutoff}$ even for huge $\omega_{\rm s2}$.

\begin{figure}[t]
\centering
\includegraphics[width=10cm]{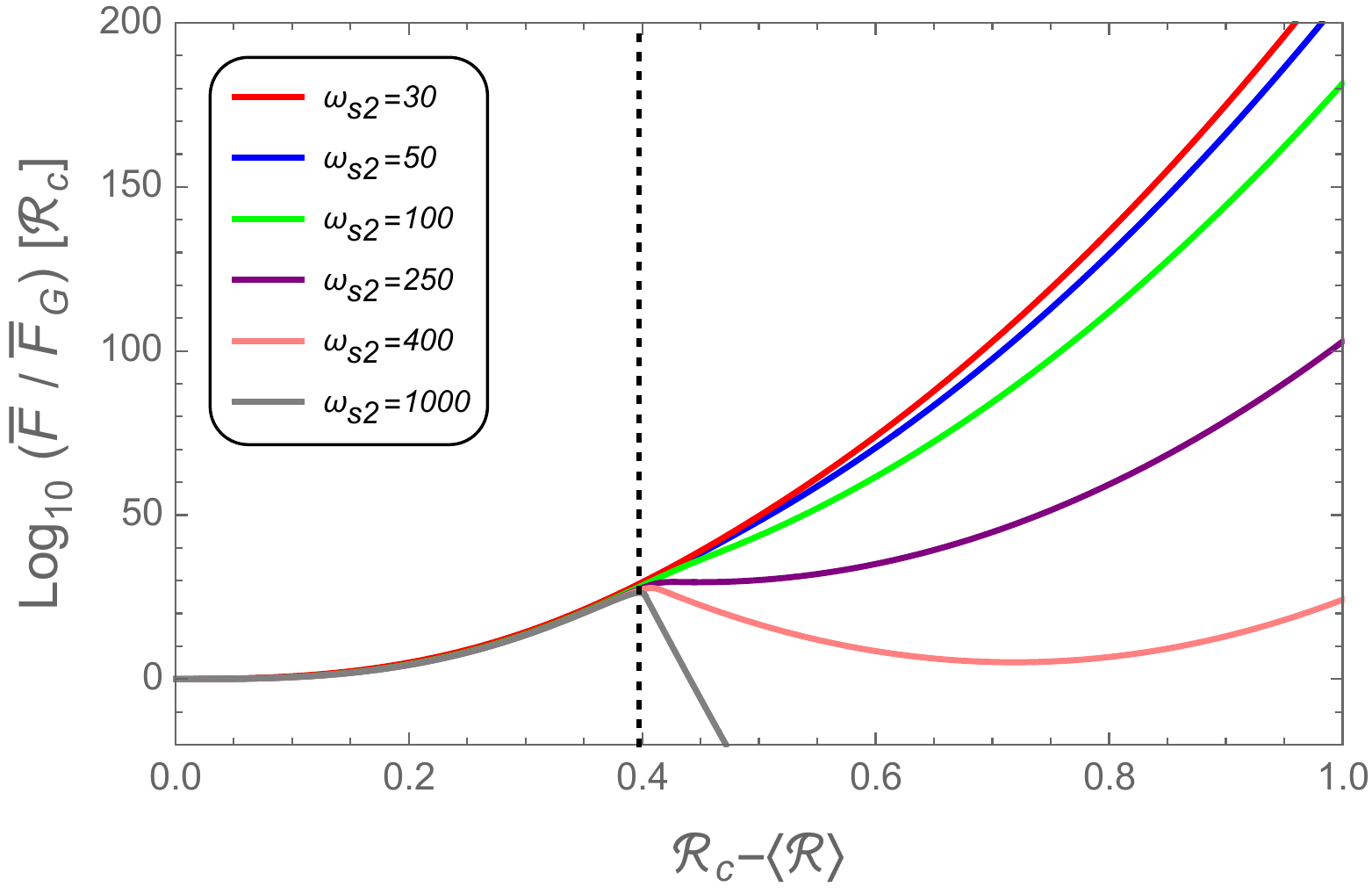}
\caption{\label{CCDF} 
The complementary cumulative distribution function (CCDF) $\bar{F}[{\cal{R}}_c]$ 
in Eq.~\eqref{bar F def} normalized by the Gaussian case $\bar{F}_G[{\cal{R}}_c]$ in Eq.~\eqref{Gauss Fbar} for $\omega_{\rm s2}=30,50,100,250,400$ and $1000$ from top to bottom.
The other parameters are the same as Fig.~\ref{PDFex}.
The vertical dotted black line represents the cutoff value ${\cal{R}}_{\rm cutoff}=\kappa g/3=0.4$.
Beyond the cutoff, the CCDF highly sensitive to $\omega_{\rm s2}\propto 1/\Delta\varphi$,
because of the exponential tail contribution.
Thus, the PBH abundance strongly depends on $\omega_{\rm s2}$, if ${\cal{R}}_c >{\cal{R}}_{\rm cutoff}$.
}
\end{figure}

With the evaluation of the PBH abundance in mind, we calculate the probability $\bar{F}[{\cal{R}}_c]$ where $\cal{R}$ is greater than or equal to the threshold value of the PBH formation ${\cal{R}}_c$, namely,
\be
\bar{F}[{\cal{R}}_c]=\int^\infty_{{\cal{R}}_c}P[{\cal{R}}]{\rm d}{\cal{R}}\,.
\label{bar F def}
\ee
It is called the complementary cumulative distribution function (CCDF) in the context of probability theory.
We also introduce the CCDF of the Gaussian PDF
\be
\bar{F}_G[{\cal{R}}_c]=\frac{1}{2}\text{Erfc}\left(\frac{{\cal{R}}_c}{\sqrt{2\sigma^2_{\cal{R}}}}\right)\,,
\label{Gauss Fbar}
\ee
where $\text{Erfc}(x)$ is the complementary error function and the variance is fixed by that of the simplified Gaussian PDF in Eq.~\eqref{eachP}, namely $\sigma_{\cal{R}}=|\beta-\kappa\gamma/3g-\gamma/\omega_{\rm s2}g^2|\sigma_{\delta\varphi}$.
Their ratio, $\bar{F}/\bar{F}_G$, is useful for understanding how much the non-Gaussianity impacts on the abundance of PBH.

In Fig.~\ref{CCDF}, the normalized CCDFs for different $\omega_{\rm s2}$ are shown against ${\cal{R}}_c-\langle\cal{R}\rangle$.
The cutoff value ${\cal{R}}_{\rm cutoff}$ is also shown as a vertical black dotted line.
It can be seen that the CCDF dramatically changes depending on the value of $\omega_{\rm s2}$, if ${\cal{R}}_c-\langle\cal{R}\rangle$ is larger than the cutoff value.
For smaller $\omega_{\rm s2}$, the normalized CCDF becomes larger beyond the cutoff, because the contribution from the exponential tail is more significant in comparison to the Gaussian case.
For huge $\omega_{\rm s2}$, $\bar{F}$ is abruptly suppressed at the cutoff but stays finite, because of the tail contribution.
These results imply that when the critical value ${\cal{R}}_c$ of the PBH formation is larger than the cutoff ${\cal{R}}_{\rm cutoff}$, it is crucially important to take into account the finite width of the upward step for estimating the PBH abundance, because otherwise one may underestimate it.

\section{Highly asymmetric PDF}
\label{Rmineff}
%
\begin{figure}[t]
\centering
\includegraphics[width=10cm]{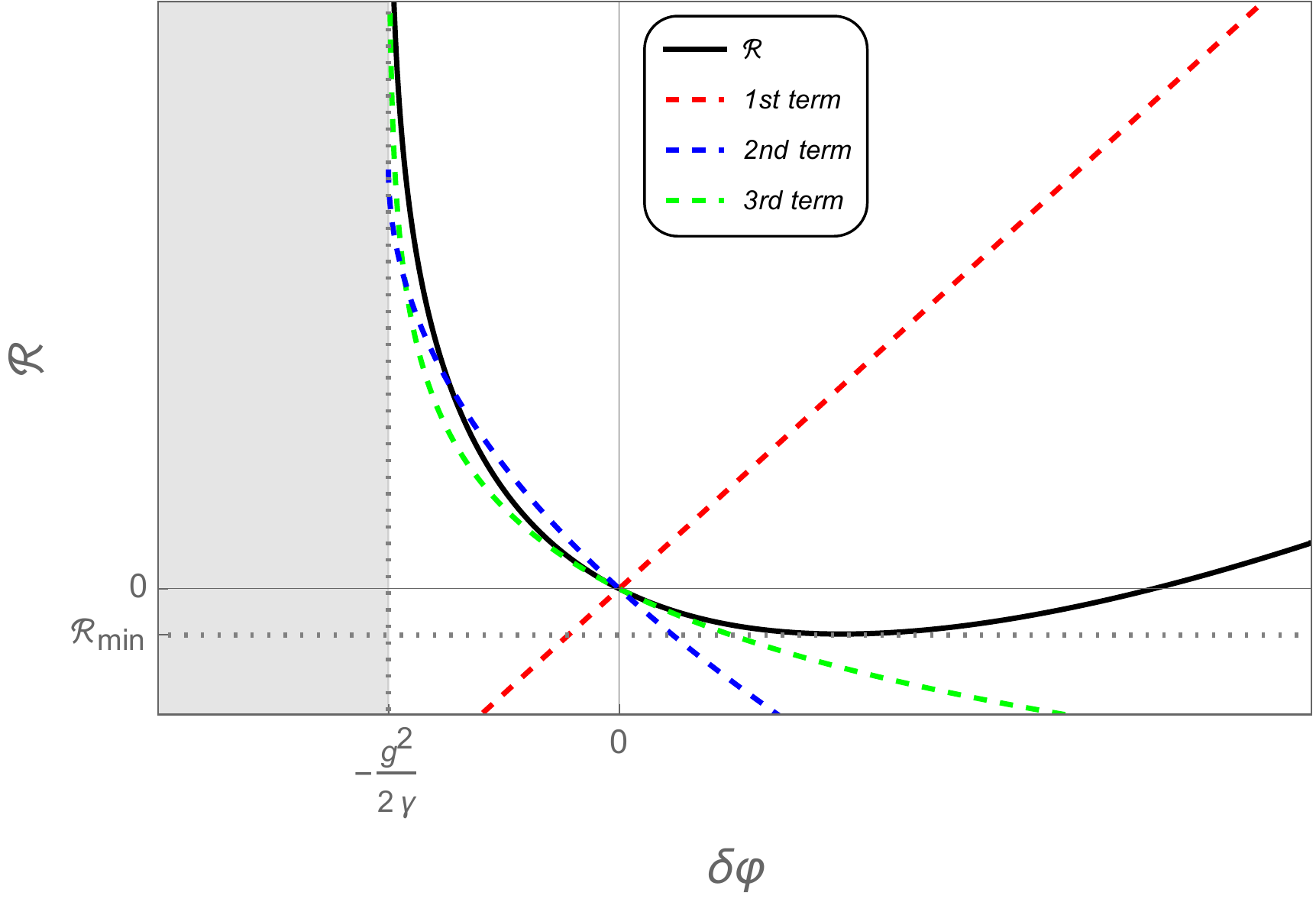}
\caption{Schematic illustrations of Eq.~(\ref{R}) for $\gamma>0$ and $\beta>\kappa\gamma/3$.
The solid black line represents the sum of the all three terms, and the dashed coloured lines represent each of them in Eq.~(\ref{R}), i.e., the first term (red), the second term (blue), and the third term (green). 
The minimum value of $\cal{R}$ is denoted by the horizontal dotted line.
The boundary, $-g^{2}/2\gamma$, is shown as the vertical dotted line.
The gray shaded region corresponds to the inflaton trajectories which fail to climb the step and are stuck at a local minimum.
\label{dphiR} }
\end{figure}
In this section we will show that the PDFs with unusual shapes can be realised at certain scales exiting the Hubble horizon before the step stage and discuss its consequences.
For the better understanding of this phenomenon, let us first revisit the relation between $\cal{R}$ and $\delta\varphi$, Eq.~\eqref{R}, which has been derived in Sec.~\ref{deltaN}.

The relation~(\ref{R}) is illustrated as a solid black line in Fig.~\ref{dphiR} 
for $\gamma>0$ and $\beta>\kappa\gamma/3$.
In the figure, the contributions of each term in Eq.~(\ref{R}) are shown as coloured dashed lines.
We can see that at around $\delta\varphi=-g^{2}/2\gamma$, the third term (green dashed line), which is the contribution from the step width, becomes dominant and causes ${\cal{R}}$ to diverge to infinity.
This gives rise to the exponential tail in the PDF of $\cal{R}$.

It should be noted that there appears a local minimum of $\cal{R}$.
For $\gamma>0$, one can see by drawing a family of trajectories in the phase space that
perturbed trajectories with $\delta\varphi<0$ have smaller velocities $|\pi|$ at $\varphi_1$ than the background trajectory (i.e. $|\pi_1+\delta\pi_1|<|\pi_1|$).
Therefore it takes an enormous amount of time for some of the trajectories to pass through the step. 
As a result, they may lead to large positive $\delta N$ for $\delta\varphi<0$.
One can also see that from Eq.~\eqref{R}, for $\beta>\kappa\gamma/3$, 
the first term dominates for sufficiently large positive $\delta\varphi$,
and $\delta N$ increases in proportion to  $\delta\varphi$.
Combining the above two facts together with the fact that 
$\delta N=0$ for $\delta\varphi=0$, 
we can conclude that a local minimum of ${\cal{R}}$ exists, which we denote by
${\cal{R}}_{\rm min}$.\footnote{ 
Note that, if $\gamma<0$, $\delta N$ is a monotonically increasing function of $\delta\varphi$ and there is no ${\cal{R}}_{\rm min}$. Therefore,
${\cal{R}}_{\rm min}$ exists only if $\gamma>0$, i.e. when the scalar field accelerates in the first SR stage.}.
In this section, we discuss the effect of the presence of ${\cal{R}}_{\rm min}$ on the PDF and the local non-Gaussianity parameter $f_{\rm NL}^{\rm local}$.

Now let us take a closer look at the PDF shape and consider its consequences.
Expanding Eq.~\eqref{R} with respect to $\delta\varphi$, we obtain 
\be
{\cal{R}}=A\delta\varphi+ B\delta\varphi^2+\mathcal{O}(\delta\varphi^3)
\label{Taylor}
\ee
with
\be
A\equiv\beta-\frac{\kappa \gamma}{3g}-\frac{\gamma}{\omega_{\rm s2}g^2}, \hspace{1cm} 
B\equiv\frac{\gamma^2}{g^2}\left[\frac{\kappa g}{6}\left(\frac{1}{g^2}-1\right)+\frac{1}{2\omega_{\rm s2}}\left(\frac{2}{g^2}-1\right)\right]\,.
\label{A and B}
\ee
With Eq.~\eqref{Taylor}, the local non-gaussianity parameter of $\cal{R}$ is defined and computed as
\be
f_{\rm NL}^{\rm local}\equiv\frac{5}{3}\frac{B}{A^2}=\frac{5}{2}\frac{\kappa g(1-g^2)+\frac{3}{\omega_{\rm s2}}(2-g^2)}{\left(\frac{3\beta g^2}{\gamma}-\kappa g-\frac{3}{\omega_{\rm s2}}\right)^2}\,.
\label{fnl2}
\ee
If $A$ vanishes (i.e. ${\cal{R}}={\cal{R}}_{\rm min}=0$ at $\delta\varphi=0$), the linear term in Eq.~\eqref{Taylor} drops, the denominator of Eq.~\eqref{fnl2} disappears, and $f_{\rm NL}^{\rm local}$ diverges
\footnote{
This corresponds to the case where $\delta N$ is always positive
for any $|\delta\varphi|$ regardless of its sign.
This means that the background trajectory passes through the step in the shortest time compared to all other perturbed trajectories.}.
It happens when $\gamma=\gamma_0\equiv3\beta g^{2}\omega_{\rm s2}/(3+\kappa g\omega_{\rm s2})>0$. Since $\gamma$ depends on $\pi/\pi_1$ (see Eq.~\eqref{gamma}), its value varies as the scale of the curvature perturbation of interest $k$ changes. 
Thus, $f_{\rm NL}^{\rm local}$ may diverge at a particular scale at which $\gamma=\gamma_0$ and $A=0$.
Although diverging $f_{\rm NL}^{\rm local}$ does not mean that any physical quantity becomes singular, it is interesting to explore what kind of dynamics occurs around this particular scale.

\begin{figure}[t]
\centering
\includegraphics[width=10cm]{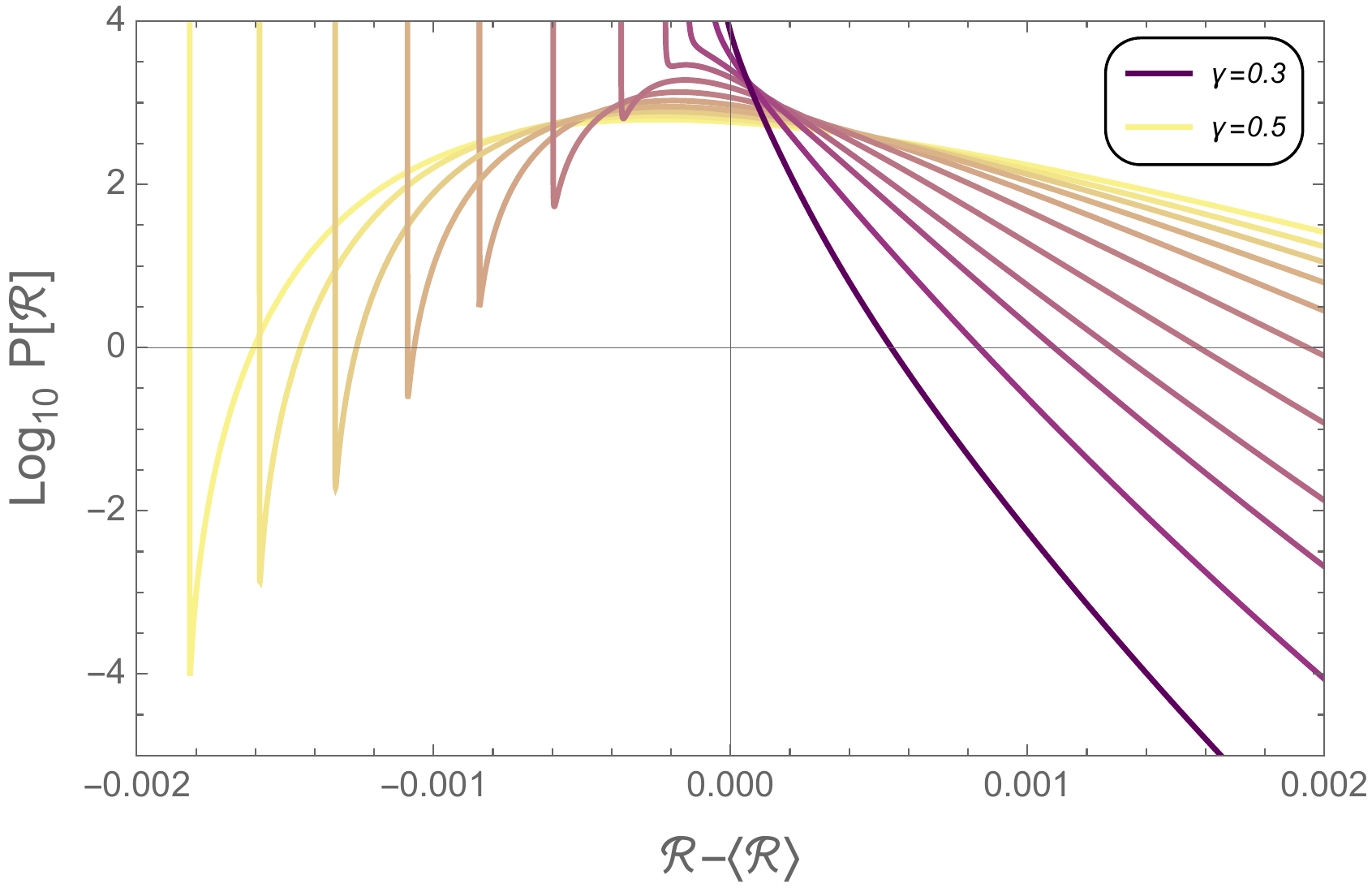}
\caption{The PDFs of $\cal{R}$ against $\cal{R}-\langle \cal{R}\rangle$ for $\kappa=1$, $\beta=20$, $g=0.02$, $\sigma_{\delta\varphi}=5\times10^{-5}$ and $\omega_{\rm s2}=50$. 
$\gamma$, which corresponds to the scale of $\cal{R}$, varies from $0.5$ (light yellow) to $\gamma_0=0.3$ (dark purple) with colour gradient. 
The probability density for negative $\cal{R}$ gradually disappears as the scale changes.
The mean value is $\langle {\cal{R}}\rangle= {\cal{O}}(10^{-5})$.
Each PDF is normalized so that the total probability is unity, and the contribution from the apparently diverging part at the lower end of the PDF is negligibly small.}
\label{PDFdiv}
\end{figure}

When $A$ vanishes, $\cal{R}$ can take only positive values and then the PDF only has its right-half part. This simple argument based on the expanded expression~\eqref{Taylor} actually applies to the fully non-linear result~\eqref{R}.
In Fig.~\ref{PDFdiv}, we present the full PDF for varying $\gamma$ from a larger value to $\gamma_0=0.3$.
One can see that the full PDF deforms from a rather symmetric shape into a completely asymmetric one as the scale changes. 
Although it is not shown there, for $\gamma<\gamma_0$ the PDF regains its left-half part.
At the lower end of the PDF of $\cal{R}$, many different values of $\delta\varphi$ corresponds to the minimum $\cal{R}$ (see Fig.~\ref{dphiR}), their probability density $P[\delta\varphi]$ condenses there, and $P[\cal{R}]$ formally diverges.
However, its contribution to the total probability is negligible and harmless.
In short, the peculiar behavior seen in Fig.~\ref{PDFdiv} is not pathological, and it is remarkable to have such a highly asymmetric PDF of $\cal{R}$. 

\begin{figure}[t]
\centering
\includegraphics[width=8.5cm]{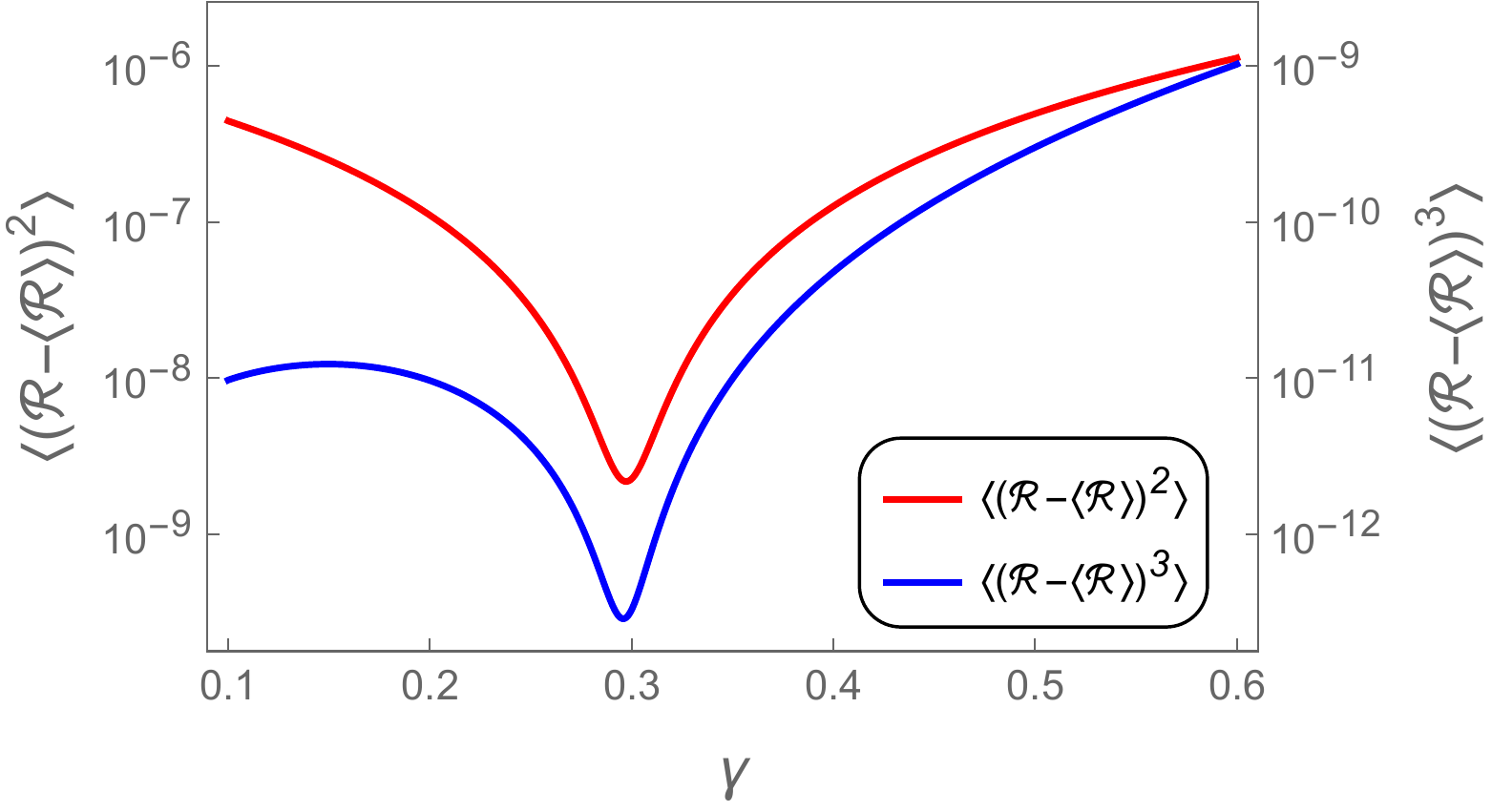}
\hspace{1cm}
\includegraphics[width=7.5cm]{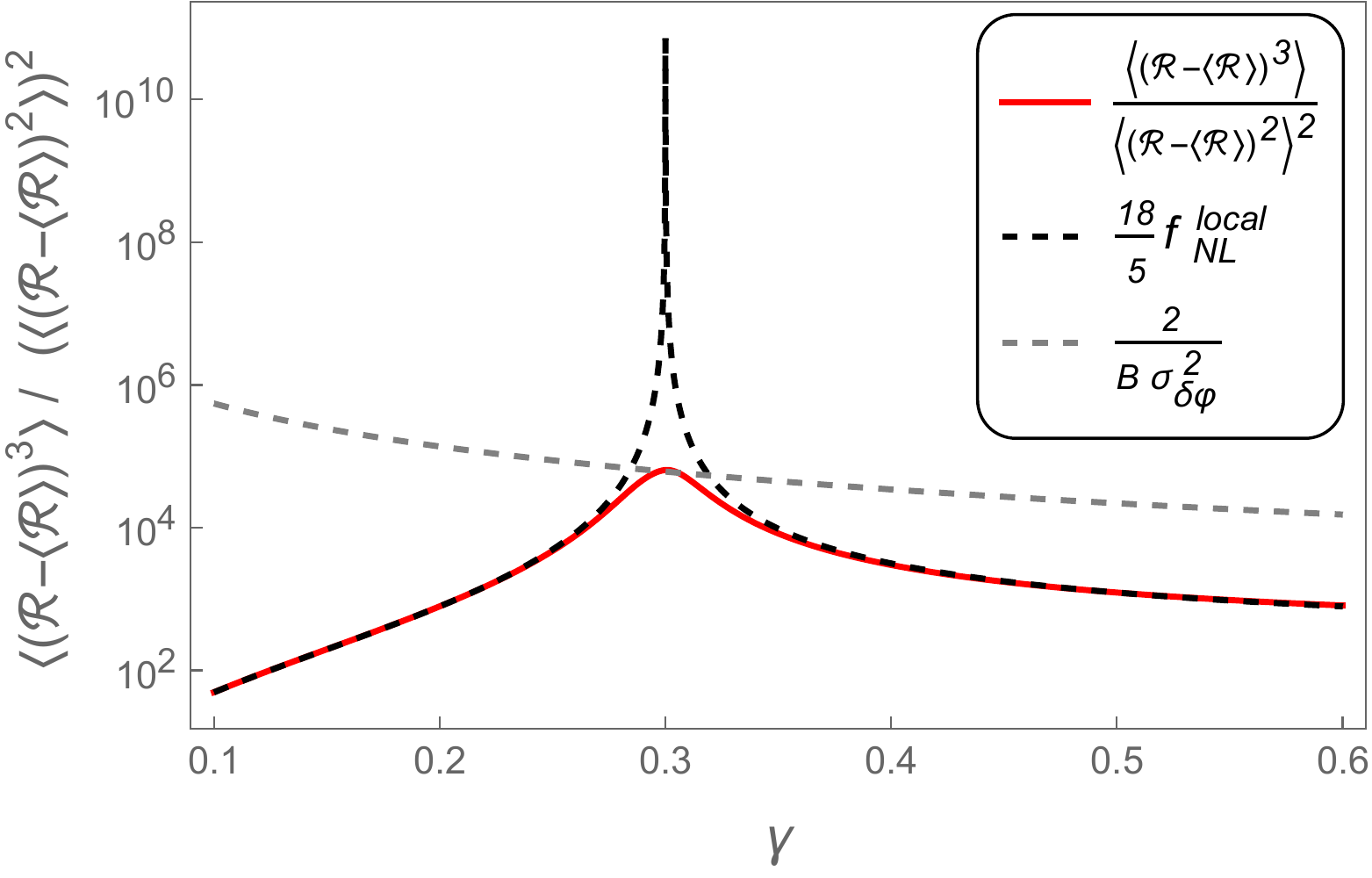}
\caption{ 
(Left panel) The variance $\langle ({\cal{R}}-\langle {\cal{R}} \rangle)^{2} \rangle$ (red) and the skewness $\langle ({\cal{R}}-\langle {\cal{R}} \rangle)^{3} \rangle$ (blue) of the curvature perturbation against $\gamma$ representing the corresponding scale around $\gamma_0=0.3$. They have sharp dips at $\gamma=\gamma_0$ because the linear term in Eq.~\eqref{Taylor} vanishes.
(Right  panel) The ratio of the these two $\langle ({\cal{R}}-\langle {\cal{R}} \rangle)^{3} \rangle/\langle ({\cal{R}}-\langle {\cal{R}} \rangle)^{2} \rangle$ compared to the asymptotic expressions, $2/B\sigma_{\delta\varphi}^2$ (gray dashed) and $18f_{\rm NL}^{\rm local}/5$ (black dashed), given in Eq.~\eqref{R3overR22}.
They match at $\gamma=\gamma_0$ and at far region from it, respectively, as expected.
The parameter choice is the same as Fig.~\ref{PDFdiv}.}
\label{R2}
\end{figure}

The most direct observables related to the PDF may be the variance and skewness of $\cal{R}$, i.e. 
$\langle ({\cal{R}}-\langle {\cal{R}} \rangle)^{2} \rangle$ and $\langle ({\cal{R}}-\langle {\cal{R}} \rangle)^{3} \rangle$.
Using Eq.~\eqref{Taylor}, their ratio can be computed as
\be
\frac{\langle ({\cal{R}}-\langle {\cal{R}} \rangle)^{3} \rangle}{\langle ({\cal{R}}-\langle {\cal{R}} \rangle)^{2} \rangle{}^2}=\frac{6A^2 B\sigma_{\delta\varphi}^4+8B^3 \sigma_{\delta\varphi}^6}{A^4 \sigma_{\delta\varphi}^4 +4A^2B^2\sigma_{\delta\varphi}^6+4B^4\sigma_{\delta\varphi}^8}
\simeq\begin{cases}\frac{18}{5}f_{\rm NL}^{\rm local} \hspace{0.2cm} (A\gg B\sigma_{\delta\varphi}) \\
\frac{2}{B\sigma_{\delta\varphi}^2} \hspace{0.65cm}  (A\ll B\sigma_{\delta\varphi}) \end{cases}
\label{R3overR22}
\ee
For $A\ll B\sigma_{\delta \varphi}$ this ratio is not related to $f_{\rm NL}^{\rm local}$, while in the other limit they coincide up to an $\mathcal{O}(1)$ numerical factor.
Fig.~\ref{R2} shows the variance, skewness and their ratio for $\gamma$ around $\gamma_0$.
Since $\gamma$ corresponds to the scale $k$, they are closely connected to the power spectrum and the bispectrum of the curvature perturbation.
Both of them has a dip at $\gamma=\gamma_0=0.3$ where the linear contribution from $\delta\varphi$ disappears (i.e. $A=0$).
This corresponds to the dip in the power spectrum that frequently appears in PBH formation models\cite{Byrnes:2018txb}.
In linear perturbation theory,
$\delta\varphi^2$ is neglected, ${\cal{R}}$ would vanish from Eq.~\eqref{Taylor}, and the dip is infinitely deep at $\gamma=\gamma_0$.\footnote{The dip would never become exactly zero in linear perturbation theory, which is in contradiction with our result. We suspect this is due to the neglection of the next leading order terms in gradient expansion in the $\delta N$ formalism. In any case, since the nonlinear terms dominate the spectrum at the dip, this inaccuracy does not affect the conclusion that the deep dip disappears once nonlinear terms are included.}
In contrast, our result with the $\delta N$ formalism 
indicates that the dip depth is finite due to the non-linear contribution.
It has been reported that the inclusion of a loop correction results in a shallower dip~\cite{Franciolini:2023lgy}, which is consistent with our result.

\begin{figure}[t]
\centering
\includegraphics[width=9.1cm]{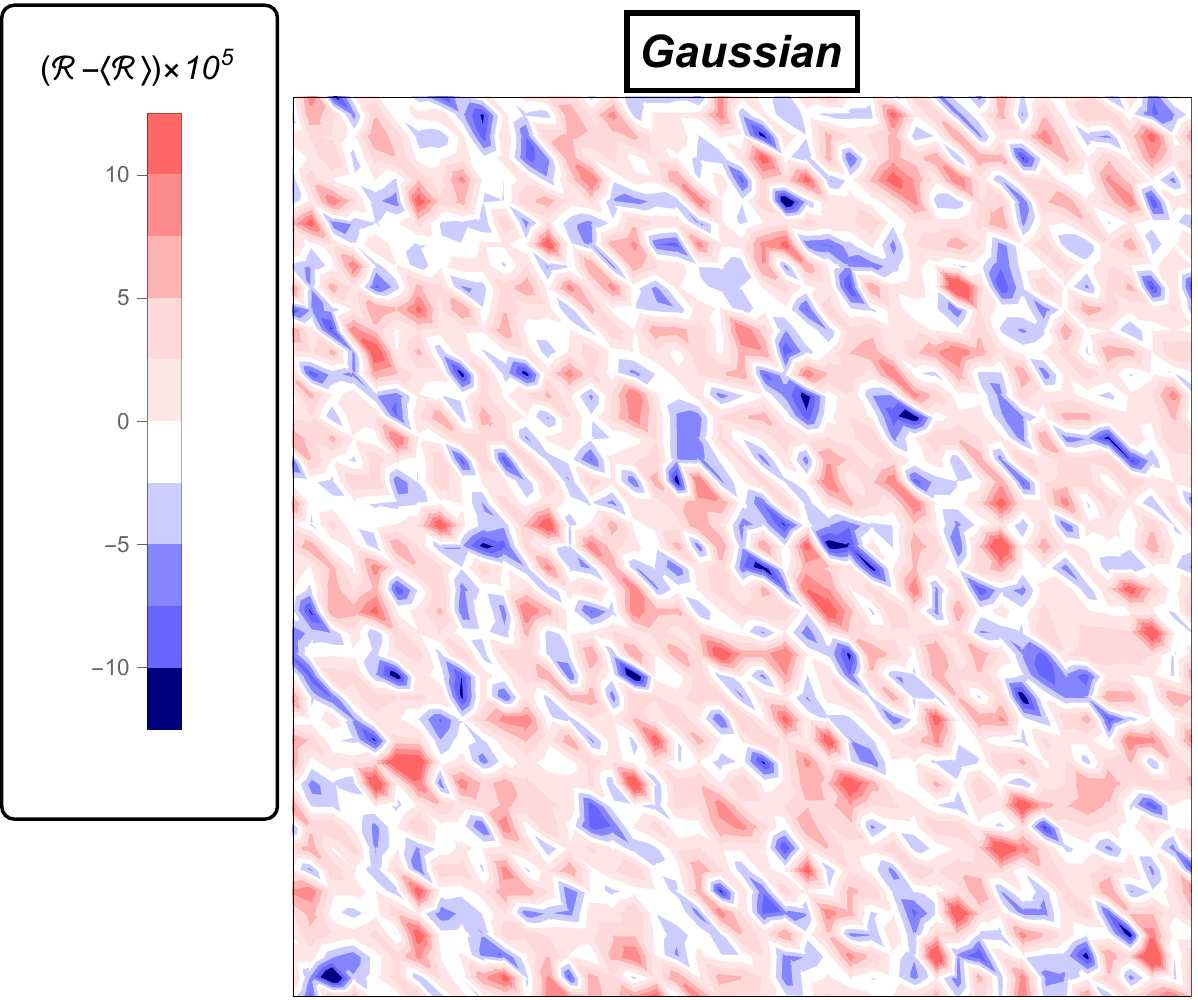}
\hspace{0.2cm}
\includegraphics[width=6.85cm]{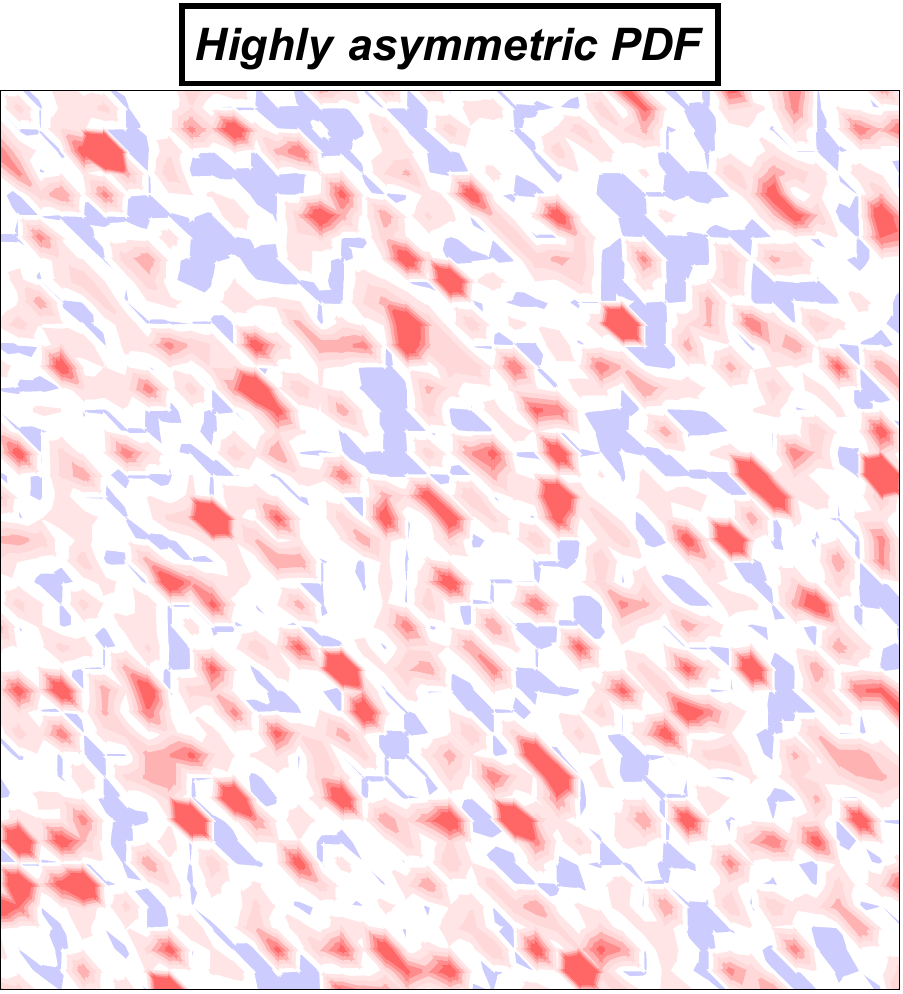}
\caption{
Illustration of the distributions of the curvature perturbation $\cal{R}$ based on different PDFs.
$\cal{R}$ is randomly assigned to each of $40 \times 40$ two-dimensional lattice points according to the Gaussian PDF (left) and the asymmetric PDF with $\gamma=\gamma_0$ (right).
Both panels use the same colour scheme, with red for high and blue for low value of $\cal{R}$.
Identifying $\cal{R}$ with the density fluctuation, one can observe the spatial distribution of the primordial density contrast. 
In the Gaussian distribution, high and low density regions appear in equal proportions, whereas very low density regions (dark blue) are not found in the highly asymmetric PDF, which implies the distribution is significantly biased.}
\label{fluc}
\end{figure}

In order to intuitively understand the consequence of the highly asymmetric PDF,
we present Fig.~\ref{fluc} in which $\cal{R}$ is regarded as a classical stochastic quantity and its values are randomly assigned to lattices on a two-dimensional plane.
$\cal{R}$ is probabilistically weighted by the Gaussian PDF in the left panel and the highly asymmetric PDF for $\gamma=\gamma_0$ in the right panel, respectively.
It can be clearly seen that the distribution is biased to positive value
in the highly asymmetric case compared to the Gaussian case.
In the highly asymmetric case, $\cal{R}$ does not appear with large negative values because of $\cal{R}_{\rm min}$.
When one interprets Fig.~\ref{fluc} as the spatial distribution of density fluctuations, this result would imply that there are fewer voids (low-density regions) than the normal Gaussian case. It would be interesting to consider such a signature in observation. We leave it for future study.

We expect that such asymmetric PDFs generically appear in models in which the inflaton experiences an accelerated SR phase, passes an inflection point, and subsequently enters a decelerated phase.
This is because when we consider $\delta N$ from the accelerated SR phase, it can become large and positive for both $\delta\varphi>0$ and $\delta\varphi<0$, which leads to the existence of ${\cal{R}}_{\rm min}$ in the same way as our model.

Assuming the inflaton $\varphi$ is initially slow-rolling and moving toward the negative direction, the scalar field perturbation $\delta\varphi$ in the accelerated phase affects the velocity at the inflection point $\pi_1$ in such a way that its absolute value $|\pi_1+\delta\pi_1|$ becomes larger for $\delta\varphi>0$. Conversely, 
 $|\pi_1+\delta\pi_1|$ becomes smaller than $|\pi_1|$ for $\delta\varphi<0$.
For such trajectories it takes longer time to pass through the decelerated phase, which leads to large positive $\delta N$.

On the other hand, for $\delta\varphi>0$, although the time to pass through the decelerated phase may become shorter by the increase in the absolute value of the velocity, depending on the values of the model parameters as well as on the scale of interest, the number of $e$-folds may eventually increase for sufficiently large $\delta\varphi$, as the perturbed trajectory starts at a greater distance from the inflection point. This gives rise to the appearance of ${\cal{R}}_{\rm min}$.
In our model the condition that this happens is $\beta>\kappa\gamma/3$.


Furthermore, corresponding to the scale at which $\gamma=\gamma_0$ in our model, we expect that there is a scale at which $\delta N$ is non-negative, regardless of the sign of $\delta\varphi$.
This occurs when the term linear in $\delta\varphi$ vanishes. Apparently, the PDF in this case is highly asymmetric.

\section{Conclusion}
\label{conclusion}
We studied an inflationary model in which the inflaton potential includes a finite-width upward step between two slow-roll stages.
The inflaton loses its kinetic energy during the step stage where it climbs up the step, which has a large effect on the statistical distribution of the curvature perturbation. We derived the relation between the curvature perturbation $\cal{R}$ and the scalar field perturbation $\delta\varphi$ and obtained Eq.~\eqref{R} by using the $\delta N$ formalism. 
We found that the step-width $\Delta\varphi$ plays a important role in the estimation of $\cal{R}$.

In Sec.~\ref{PDF}, we calculated the PDF of the curvature perturbation $\cal{R}$ and obtained Eq.~\eqref{PDFofR}.
The result is shown in Fig~\ref{PDFex}.
For ${\cal{R}}<{\cal{R}}_{\rm cutoff}=\kappa g/3$ the PDF follows the Cutoff PDF given in Eq.~\eqref{eachP}, while for ${\cal{R}}>{\cal{R}}_{\rm cutoff}=\kappa g/3$ the exponential tail $P[{\cal{R}}]\propto\exp(-2\omega_{\rm s2}{\cal{R}})$ is dominant. 
The slope of the tail depends on the step width $\Delta\varphi$ through $\omega_{\rm s2}\simeq \sqrt{2}~ |\pi_1|/\Delta\varphi$.
The CCDF was also calculated, and the significant impact on the PBH abundance of this exponential tail was illustrated in Fig.~\ref{CCDF}.
We conclude that the step width should be taken into account for the accurate estimation of the PBH abundance.

In Sec.~\ref{Rmineff} we discussed the highly asymmetric PDF.
As shown in Fig.~\ref{dphiR}, for $\gamma>0$ and $\beta>\kappa\gamma/3$, $\cal{R}$ has a minimum value ${\cal{R}}_{\rm min}$.
At a particular scale $\gamma=\gamma_0$ exiting the Hubble horizon before the step, $\cal{R}_{\rm min}$ is realised with $\delta\varphi=0$, which means that $A$ as defined in Eq.~\eqref{A and B} vanishes and that $\cal{R}$ is supported by $\delta\varphi^2$ instead of $\delta\varphi$.
We found that this scale $\gamma_0$ corresponds to the dip of the curvature power spectrum known in other models. 
On this scale, $f_{\rm NL}^{\rm local}$ formally diverges and the highly asymmetric PDF appears as shown in Figs.~\ref{PDFdiv} and \ref{R2}.
The highly asymmetric PDF may result in a lower abundance of low-density regions such as voids (see Fig.~\ref{fluc}) compared to the normal Gaussian case.

In this paper, we mainly investigated curvature perturbations on the scales exiting the Hubble horizon just before the background inflaton enters the step stage. 
It would also be interesting to calculate the PDF for scales exiting the Hubble horizon during and after the step stage for the comprehensive understanding of this model.
In doing so, one needs to track the time evolution of $\delta\varphi$ more precisely, to include the contribution of $\delta\pi$, and to take into account the intrinsic non-gaussianity of $\delta\varphi$.
As another approach, it is also fascinating to apply the stochastic-$\delta N$ formalism to this upward step model.
We leave these as future work.

\section*{Acknowledgments}
We thank Shinji Tsujikawa for useful discussions and comments. 
This work is supported by JSPS KAKENHI grants 18K13537, 20H05854, 23K03424 (TF), 20H05853 (MS), and by the World Premier International Research Center Initiative (WPI Initiative), MEXT, Japan.


\renewcommand{\theequation}{A.\arabic{equation}}
\setcounter{equation}{0}

\appendix

\section{Parameters in {\boldmath$f_{\rm step}$}}
\label{paraofstep}
The parameters in Eq.~\eqref{step} are determined by imposing the continuity conditions on the potential $v(\varphi)$ and its first derivative $\partial_\varphi v$ at $\varphi=\varphi_1, \varphi_2$ and $\varphi_c$.
We then obtain the parameters of $f_{\rm step}$ as 
\ba
& &
A_1=v_{\rm sr1}(\varphi_{1})-\frac{\left(\partial_{\varphi}v_{\rm sr1}(\varphi_1)\right)^2}{4B_1}
\,, \label{A1}\\
& &
A_2=v_{\rm sr2}(\varphi_{2})-\frac{\left(\partial_{\varphi}v_{\rm sr2}(\varphi_2)\right)^2}{4B_2}
\,, \\
& &
B_1=\frac{2\Delta v}{(\Delta\varphi)^2}+\frac{3\partial_{\varphi}v_{\rm sr1}(\varphi_1)+\partial_{\varphi}v_{\rm sr2}(\varphi_2)}{2\Delta\varphi}\,, \\
& &
B_2=-\frac{2\Delta v}{(\Delta\varphi)^2}-\frac{\partial_{\varphi}v_{\rm sr1}(\varphi_1)+3\partial_{\varphi}v_{\rm sr2}(\varphi_2)}{2\Delta\varphi}\,, \\
& &
\varphi_{\rm min}=\varphi_1-\frac{\partial_{\varphi}v_{\rm sr1}(\varphi_1)}{2B_1}\,, \\
& &
\varphi_{\rm max}=\varphi_2-\frac{\partial_{\varphi}v_{\rm sr2}(\varphi_2)}{2B_2},
\label{appconstants}
\ea
where $\Delta v\equiv v_{\rm sr2}(\varphi_2)-v_{\rm sr1}(\varphi_1)$ corresponds to the height of the step.

\section{Solving background EOM}
\label{solvingBGEOM}
\renewcommand{\theequation}{B.\arabic{equation}}
\setcounter{equation}{0}
Here we solve the background equation of motion for the scalar field.
In the following, we assume that the kinetic energy of the scalar field is subdominant compared to the potential energy, $h^2\pi^2\ll v$, throughout the three stages (approximation (I) in Sec.~\ref{Background}).
This reduces the EOM (\ref{fieldeq}) to 
\be
\frac{{\rm d}\pi}{{\rm d}n}+3\pi+3\frac{\partial_{\varphi}v}{v}=0\,.
\label{appfieldeq}
\ee
We analytically solve this approximate EOM (\ref{appfieldeq}) in each of the three stages.

\subsection{SR stages}
\label{SRstages}
In this subsection we discuss the dynamics before and after the step. The step stage will be addressed in the next subsections.
In these stages, the potential is almost flat and the Taylor expansion is a sufficiently good approximation for the potential. 
To solve Eq.~(\ref{appfieldeq}) analytically, we expand the potential term up to the first order in the vicinity of $\varphi=\varphi_i$.
Then, we obtain the equation of motion
\be
\frac{{\rm d}\pi}{{\rm d}n}+3\pi+3\sqrt{2\epsilon_{Vi}}-\frac{3}{2}\eta_{i}(\varphi-\varphi_i)=0\,,
\label{appBGeqSR}
\ee
for $\varphi\ge\varphi_1$ ($i=1$) and $\varphi\le\varphi_2$ ($i=2$), respectively.
The solutions are 
\be
\varphi(n)-\varphi_i=C_{i}^+(\pi_i) e^{\lambda_{i}^{+} (n-n_i)}+C_{i}^-(\pi_i) e^{\lambda_{i}^{-} (n-n_i)}+\frac{2\sqrt{2\epsilon_{Vi}}}{\eta_{i}}\,,
\label{appBGsolSR}
\ee
where $\lambda_{i}^{\pm}$ and $C_{i}^{\pm}$ are constants for a given trajectory (i.e. for given boundary conditions at $n=n_i$),
\ba
& &
\lambda_{i}^{+}=-\frac{3}{2}\left(1+\sqrt{1+\frac{2\eta_{i}}{3}}\right)
\simeq-3-\frac{\eta_i}{2}
\,, \\
& &
\lambda_{i}^{-}=-\frac{3}{2}\left(1-\sqrt{1+\frac{2\eta_i}{3}}\right)
\simeq\frac{\eta_i}{2}
\,, \\
& &
C_i^+(\pi_i)=\frac{1}{\lambda_{i}^{+}-\lambda_{i}^{-}}\left(\pi_i+\lambda_{i}^{-}\frac{2\sqrt{2\epsilon_{Vi}}}{\eta_{i}}\right)
\simeq-\frac{1}{3}\left(\pi_i+\sqrt{2\epsilon_{Vi}}\right)
\,,\label{appCp}\\
& &
C_i^-(\pi_i)=\frac{1}{\lambda_{i}^{-}-\lambda_{i}^{+}}\left(\pi_i+\lambda_{i}^{+}\frac{2\sqrt{2\epsilon_{Vi}}}{\eta_{i}}\right)
\simeq\frac{1}{3}\left(-\frac{6\sqrt{2\epsilon_{Vi}}}{\eta_{i}}+\pi_i+\sqrt{2\epsilon_{Vi}}\right)
\,.
\label{Cm}
\ea

For later discussion, we show $\pi_2$ with the first-order $\Delta\varphi$-corrections from the effect of the Hubble friction term.
From the energy conservation law which includes the dissipation caused by the Hubble friction term, we obtain 
\ba
\pi_2&=&-\sqrt{\pi_1^2-6\log\left(\frac{v(\varphi_2)}{v(\varphi_1)}\right)-6\int^{n_2}_{n_1}\pi^2(n){\rm d}n}\,\label{apppi2}\\
&\simeq&-\sqrt{\pi_1^2-6\log\left(\frac{v(\varphi_2)}{v(\varphi_1)}\right)-3\pi_1^2 N^{(\rm s1)}+3\pi_c\Delta\varphi-3\pi_2^2 N^{(\rm s2)}}\,,
\label{apppi22}
\ea
where we used the zeroth-order solutions \eqref{BGsolstep1} and \eqref{BGsolstep2} to derive the second line. 
In Eq.~(\ref{apppi2}), the third term inside the square root corresponds to the dissipation term, which we ignored in the main context as the approximation (I\hspace{-1.2pt}I\hspace{-1.2pt}I) in Sec.~\ref{Background}.

Note, however, that the last three terms in Eq.~\eqref{apppi22} are not negligible compared to the sum of the first two terms for $\omega_{\rm s1}g^2<1$. 
Therefore, for the purpose of estimating $\pi_2$, the approximation (I\hspace{-1.2pt}I\hspace{-1.2pt}I) is not a good approximation.
However, as we will see later in Appendix.~\ref{Hubblefriction}, the approximation (I\hspace{-1.2pt}I\hspace{-1.2pt}I) can be considered as a good approximation for estimating $\delta\pi_2$.

\subsection{S1 region}
\label{appS1region}
Substituting Eq.~(\ref{step}) into Eq.~(\ref{appfieldeq}) yields the equation of motion to be solved.
To solve it analytically, we approximate the denominator of the third term in Eq.~(\ref{appfieldeq}) to be the constant value $A_1$ shown in Eq.~(\ref{A1}).
This approximation is valid under the assumption that the step height $\Delta v$ is sufficiently smaller than the value of the potential.
Then the EOM is given by
\be
\frac{{\rm d}\pi}{{\rm d}n}+3\pi+\omega_{\rm s1}^2(\varphi-\varphi_{\rm min})=0\,.
\label{appBGeqstep1}
\ee
Assuming $\omega_{\rm s1}^2>9/4$ and setting the values at $n=n_1$ as $\varphi=\varphi_1$ and $\pi=\pi_1$, the solution is
\be
\varphi(n)-\varphi_{\rm min}=D_{\rm s1} e^{-\frac{3}{2}(n-n_1)}\sin\left(\sqrt{\omega_{\rm s1}^2-\frac{9}{4}}(n-n_1)+\theta_{\rm s1}\right)\,,
\label{appBGsolstep1}
\ee
where $D_{\rm s1}$ and $\theta_{\rm s1}$ are constants for a given trajectory, defined by
\ba
& &
D_{\rm s1}= -\sqrt{\frac{\pi_1^2+3\pi_1(\varphi_1-\varphi_{\rm min})+\omega_{\rm s1}^2(\varphi_1-\varphi_{\rm min})^2}{\omega_{\rm s1}^2-\frac{9}{4}}}
\,, \\
& &
\cos\theta_{\rm s1}=\frac{\pi_1+\frac{3}{2}(\varphi_1-\varphi_{\rm min})}{D_{\rm s1}\sqrt{\omega_{\rm s1}^2-\frac{9}{4}}}\hspace{0.5cm} \text{and}\hspace{0.5cm} \sin\theta_{\rm s1}=\frac{\varphi_1-\varphi_{\rm min}}{D_{\rm s1}}
\,.
\label{appCstep1}
\ea

Here we check that if $\Delta\varphi$ is sufficiently small, Eq.~\eqref{appBGsolstep1} coincides with the solution, Eq.~\eqref{BGsolstep1}, obtained from the approximated EOM to which approximations (I\hspace{-1.2pt}I) and (I\hspace{-1.2pt}I\hspace{-1.2pt}I) are applied.
For this purpose, it is sufficient to compare $\omega_{\rm s1}^2(\varphi_1-\varphi_{\rm min})^2$ to $\pi_1^2$,
\be
\frac{\omega_{\rm s1}^2(\varphi_1-\varphi_{\rm min})^2}{\pi_1^2}=\frac{3(\partial_{\varphi}v_{\rm sr1}(\varphi_1))^2 }{2A_1 B_1 \pi_1^2}\simeq\frac{3}{4}\frac{(\partial_{\varphi}v_{\rm sr1}(\varphi_1))^2}{v_{\rm sr1}(\varphi_1)}\frac{(\Delta\varphi)^2}{ \pi_1^2\Delta v}\sim9\epsilon_{V1}\frac{(\Delta\varphi)^2}{\pi_1^4}={\cal{O}}\left(\frac{1}{\omega_{\rm s1}^2}\right)\,,\label{omegas1app}
\ee
where we used the approximation $\Delta v\sim v_{\rm sr1}(\varphi_1)\pi_1^2/6$, which can be derived by the energy conservation law~(\ref{pi2}).
From Eq.~\eqref{omegas1app}, it follows that if $\omega_{\rm s1}$ is much larger than 1, $D_{\rm s1}$, $\theta_{\rm s1}$ and $\varphi_{\rm min}$ are reduced to $\pi_1/\omega_{\rm s1}$, 0 and $\varphi_1$, respectively, i.e. Eq.~\eqref{BGsolstep1} is reproduced from Eq.~\eqref{appBGsolstep1}.

\subsection{S2 region}
\label{appS2region}
Similarly, for the S2 region, the EOM to be solved is
\be
\frac{{\rm d}\pi}{{\rm d}n}+3\pi-\omega_{\rm s2}^2(\varphi-\varphi_{\rm max})=0\,,
\label{appBGeqstep2}
\ee
where we used the fact that $B_2$ is a negative value.
Setting the values at $n=n_2$ as $\varphi=\varphi_2$ and $\pi=\pi_2$, the solution is
\be
\varphi(n)-\varphi_{\rm max}=D_{\rm s2} e^{-\frac{3}{2}(n-n_2)}\sinh\left(\sqrt{\omega_{\rm s2}^2+\frac{9}{4}}(n-n_2)+\theta_{\rm s2}\right)\,,
\label{appBGsolstep2}
\ee
where $D_{\rm s2}$ and $\theta_{\rm s2}$ are constants for a given trajectory, defined by
\ba
& &
D_{\rm s2}= -\sqrt{\frac{\pi_2^2+3\pi_2(\varphi_2-\varphi_{\rm max})-\omega_{\rm s2}^2(\varphi_{2}-\varphi_{\rm max})^2}{\omega_{\rm s2}^2+\frac{9}{4}}}
\,, \\
& &
\theta_{\rm s2}=\log\left[\frac{1}{D_{\rm s2}\sqrt{\omega_{\rm s2}^2+\frac{9}{4}}}\left(\pi_2+\left(\frac{3}{2}+\sqrt{\omega_{\rm s2}^2+\frac{9}{4}}\right)(\varphi_2-\varphi_{\rm max})\right)\right]
\,.
\label{appCstep2}
\ea

Similar to Eq.~\eqref{omegas1app}, we compare $\omega_{\rm s2}^2(\varphi_{2}-\varphi_{\rm max})^2$ to $\pi_2^2$ to find a condition under which Eq.~\eqref{BGsolstep2} is reproduced.
We obtain
\be
\frac{\omega_{\rm s2}^2(\varphi_2-\varphi_{\rm max})^2}{\pi_2^2}\simeq\frac{3}{4}\frac{(\partial_{\varphi}v_{\rm sr2}(\varphi_2))^2}{v_{\rm sr2}(\varphi_2)}\frac{(\Delta\varphi)^2}{\pi_2^2\Delta v}\sim9\epsilon_{V2}\frac{(\Delta\varphi)^2}{\pi_1^2\pi_2^2}={\cal{O}}\left(\frac{1}{\kappa^2 g^2 \omega_{\rm s2}^2}\right)\,,\label{omegas2app}
\ee
where we used the approximation $\Delta v\sim v_{\rm sr2}(\varphi_2)\pi_1^2/6$.
Thus, if $\kappa g\omega_{\rm s2}\gg1$ is satisfied, then $D_{\rm s2}$, $\theta_{\rm s2}$ and $\varphi_{\rm max}$ can be approximated by $\pi_2/\omega_{\rm s2}$, 0 and $\varphi_2$ and Eq.~\eqref{appBGsolstep2} reduces to Eq.~\eqref{BGsolstep2}.

\section{Derivation of {\boldmath$\delta N$} in each stage}
\label{appdeltaN}
\renewcommand{\theequation}{C.\arabic{equation}}
\setcounter{equation}{0}

\subsection{First SR stage}
\label{apppSR1}
In the first SR stage the background trajectory is on the SR attractor.
From Eq.~(\ref{appBGsolSR}), we have
\ba
& &
\varphi-\varphi_1=-\frac{2\sqrt{2\epsilon_{V1}}}{\eta_{1}}e^{-\lambda_1^- N^{(1)}}+\frac{2\sqrt{2\epsilon_{V1}}}{\eta_{1}}\,,
\label{appBGsol1phi}\\
& &
\pi=-\sqrt{2\epsilon_{V1}}+\frac{\eta_1}{2}(\varphi-\varphi_1)\,,
\label{appBGsol1pi}
\ea
where $N^{(1)}\equiv N(\varphi,\pi;\varphi_1,\pi_1)$.

Now, what we are here to discuss is the deviation from the SR attractor, thus it is not sufficient just to perturb the background solutions (\ref{appBGsol1phi}) and (\ref{appBGsol1pi}). 
Replacing $\pi_1$ with $\pi_1+\delta\pi_1$ in $C_1^{\pm}$, Eq.\eqref{appBGsolSR} gives a perturbed trajectory,
\ba
& &
\varphi(n)-\varphi_1=C_{1}^+(\pi_1+\delta\pi_1) e^{\lambda_{1}^{+} (n-n_1')}+C_1^-(\pi_1+\delta\pi_1) e^{\lambda_{1}^{-} (n-n_1')}+\frac{2\sqrt{2\epsilon_{V1}}}{\eta_{1}}\,,
\label{appPertphisol1}\\
& &
\pi(n)=C_{1}^+(\pi_1+\delta\pi_1)\lambda_{1}^{+} e^{\lambda_{1}^{+} (n-n_1')}+C_1^-(\pi_1+\delta\pi_1) \lambda_{1}^{-} e^{\lambda_{1}^{-} (n-n_1')}\,,
\label{appPertpisol1}
\ea
which correspond to a trajectory passing through ($\varphi_1$, $\pi_1+\delta\pi_1$) at $n=n_1'$.
We assign $\delta\varphi$ so that, in the phase space, ($\varphi+\delta\varphi$, $\pi$) lies on the perturbed trajectory given by Eqs.~(\ref{appPertphisol1}) and (\ref{appPertpisol1}).
Note that $\delta\varphi$ is uniquely determined by $\delta\pi_1$. 
Conversely, $\delta\pi_1$ should not be chosen such that there is no corresponding $\delta\varphi$.
This allows us to substitute $\varphi(n')=\varphi+\delta\varphi$ and $\pi(n')=\pi$ into Eqs.~(\ref{appPertphisol1}) and (\ref{appPertpisol1}) as values at $n=n'$.
Then, solving the system of equations for $e^{\lambda_1^{+}(n'-n_1')}$ and $e^{\lambda_1^{-}(n'-n_1')}$, we obtain
\ba
& &
\delta\pi_1\simeq-\lambda_1^{-}\delta\varphi e^{\lambda_{1}^{+}{\cal{N}}^{(1)}} 
\,,\label{appPertSolp}\\
& &
\delta\pi_1
\simeq
\left(-\lambda_1^{+}\delta\varphi+\left(1-\frac{\lambda_1^{+}}{\lambda_1^{-}}\right)\pi_*\right)e^{\lambda_1^{-}{\cal{N}}^{(1)}}-\left(1-\frac{\lambda_1^{+}}{\lambda_1^{-}}\right)\pi_1
\,,\label{appPertSolm}
\ea
where ${\cal{N}}^{(1)}\equiv N(\varphi+\delta\varphi,\pi;\varphi_1,\pi_1+\delta\pi_1)=n_1'-n'$.
To derive these results, we used the background solution (\ref{appBGsol1pi}) and $\pi_1=-\sqrt{2\epsilon_{V1}}$.

What we need to know in particular is the relation to $\delta\varphi$ for each of $\delta\pi_1$ and $\delta N^{(1)}$($\equiv {\cal{N}}^{(1)}-N^{(1)}$).
In principle, it is possible to solve them because there are two equations for two unknown quantities.
However, since this system of equations contains two exponential functions, it is not possible to obtain an explicit form of the analytic solution.
In order to solve them analytically, we limit our discussion to considering only the case in which $\delta N^{(1)}$ is sufficiently small, and perform an approximation that ignores second and higher orders for $\delta N^{(1)}$ and $\delta\varphi$.
That is, the exponential functions in Eqs.~(\ref{appPertSolp}) and (\ref{appPertSolm}) are approximately
\be
e^{\lambda_1^{\pm} {\cal{N}}^{(1)}}=e^{\lambda_1^{\pm} N^{(1)}}e^{\lambda_1^{\pm} \delta N^{(1)}}
\simeq \left(\frac{\pi_1}{\pi}\right)^{\frac{\lambda_1^{\pm}}{\lambda_1^{-}}}\left(1+\lambda_1^{\pm}\delta N^{(1)}\right)\,,
\label{expexp}
\ee
where we used $e^{\lambda_1^{-} N^{(1)}}\simeq \pi_1/\pi$, which is derived from Eqs.~(\ref{appBGsol1phi}) and (\ref{appBGsol1pi}).
Notice that this approximation~(\ref{expexp}) is only valid if $|\delta N^{(1)}|\ll 1/3$ is satisfied.
This then makes it possible to solve the system of equations analytically, and the final results can be obtained as
\ba
& &
\delta\pi_1\simeq -\lambda_1^{-}\left(\frac{\pi_1}{\pi}\right)^{\frac{\lambda_1^{+}}{\lambda_1^{-}}}\delta\varphi
\simeq -\frac{\eta_1}{2}\left(\frac{\pi}{\pi_1}\right)^{\frac{6}{\eta_1}}\delta\varphi\,,
\label{deltapi1}\\
& &
\delta N^{(1)}\simeq\frac{\lambda_1^{+}-\lambda_1^{-}\left(\frac{\pi}{\pi_1}\right)^{1-\frac{\lambda_1^{+}}{\lambda_1^{-}}}}{\lambda_1^{-}-\lambda_1^{+}}\frac{\delta\varphi}{\pi}
\simeq -\frac{\delta\varphi}{\pi} \,.
\ea
Note that in the above discussion $\delta N^{(1)}$ is limited to being small, but $N^{(1)}$ is not limited at all.

\subsection{Second SR stage}
\label{pRelaxation}

By solving Eq.~(\ref{appBGsolSR}) for $\varphi\le\varphi_2$ in conjunction with the equation for $\pi$ derived by differentiating Eq.~(\ref{appBGsolSR}) with respect to $n$, we obtain $N^{(2)}(\equiv N(\varphi_2,\pi_2;\varphi_f,\pi_f)= n_f-n_2)$ as
\be
N^{(2)}\simeq \frac{1}{\lambda_2^{-}}\log \left(\frac{\left(1-\frac{\lambda_2^{+}}{\lambda_2^{-}}\right)\pi_f}{\pi_2+\frac{\lambda_2^{+}}{\lambda_2^{-}}\sqrt{2\epsilon_{V2}}}\right)\,,
\ee
where $\pi_f$ is the value of $\pi$ at $n=n_f$.
Replacing $\pi_2$ with $\pi_2+\delta\pi_2$ in $C_2^{\pm}$, Eq.\eqref{appBGsolSR} gives a perturbed trajectory.
We can immediately obtain 
\be
{\cal{N}}^{(2)}\simeq \frac{1}{\lambda_2^{-}}\log \left(\frac{\left(1-\frac{\lambda_2^{+}}{\lambda_2^{-}}\right)(\pi_f+\delta\pi_f)}{\pi_2+\delta\pi_2+\frac{\lambda_2^{+}}{\lambda_2^{-}}\sqrt{2\epsilon_{V2}}}\right)\,,
\ee
where ${\cal{N}}^{(2)}\equiv N(\varphi_2,\pi_2+\delta\pi_2;\varphi_f,\pi_f)$.
Recall that we ignore $\delta\pi_f$ by choosing $n_f$ as the time later than which the trajectory converges to the SR attractor.
Then $\delta N^{(2)}$ is given by 
\be
\delta N^{(2)}\equiv{\cal{N}}^{(2)}-N^{(2)}\simeq-\frac{1}{\lambda_2^{-}}\log\left(1+\frac{\frac{\delta\pi_2}{\pi_2}}{1-\frac{\lambda_2^{+}}{\lambda_2^{-}}\frac{1}{\kappa g}}\right)
\simeq
-\frac{\kappa g}{3}\frac{\delta\pi_2}{\pi_2}\,.
\label{deltaN2}
\ee
Note that the last approximate equality in Eq.~(\ref{deltaN2}) is an approximation due to the fact that $|\lambda_2^{+}/\lambda_2^{-}|$ is much larger than 1, not that $\delta\pi_2/\pi_2$ is small. 
Substituting Eq.~\eqref{deltapi2} into Eq.~\eqref{deltaN2} leads to Eq.~\eqref{deltaN22}, but note that Eq.~\eqref{deltapi2} ignores the Hubble friction.
In Appendix.~\ref{Hubblefriction}, we will discuss the validity of Eq.~\eqref{deltapi2}.

\subsection{Step stage}
\label{pStep}
First, let us estimate the number of e-folds in the S1 region.
Assuming that $\Delta\varphi$ is sufficiently small, i.e. $\omega_{\rm s1}\gg1$, Eq.~(\ref{appBGsolstep1}) gives 
\be
-\frac{1}{2}\Delta\varphi\simeq\frac{\pi_1}{\omega_{\rm s1}}e^{-\frac{3}{2}N^{(\rm s1)}}\sin{\left(\omega_{\rm s1}N^{(\rm s1)}\right)}\,,
\ee
where $N^{(\rm s1)}\equiv N(\varphi_1,\pi_1;\varphi_c,\pi_c)= n_c-n_1$.
The scalar field should not oscillate around the local minimum of the potential as it climbs through the step.
This fact tells us that
\be
N^{(\rm s1)}\le\frac{\pi}{2\omega_{\rm s1}}\ll 1\,,
\ee
where we used that $\omega_{\rm s1}$ is an inverse power of the order of $\Delta\varphi$ and is very large compared to 1.
In this context, since the damping factor $\exp(-3N^{(\rm s1)}/2)$ is approximately equal to 1, we have
\be
N^{(\rm s1)}\simeq\frac{1}{\omega_{\rm s1}}\arcsin\left(-\frac{\Delta\varphi}{2\pi_1}\omega_{\rm s1}\right)\,.
\label{appNs1}
\ee
Then varying $N^{(\rm s1)}$ with respect to $\delta\pi_1$ yields
\be
\delta N^{(\rm s1)}=-\frac{1}{2}\frac{\Delta\varphi}{\sqrt{\pi_1^2-\left(\frac{1}{2}\Delta\varphi\omega_{\rm s1}\right)^2}}\frac{\delta\pi_1}{\pi_1}
\simeq
-\frac{1}{\omega_{\rm s1}}\frac{\delta\pi_1}{\pi_1}
\simeq
-\frac{\eta_1}{2\omega_{\rm s1}}\left(\frac{\pi}{\pi_1}\right)^{1+\frac{6}{\eta_1}}\delta N^{(1)}
\label{deltaNs1}
\,.
\ee
Here, $\delta\pi_1/\pi_1$ is assumed to be much smaller than 1, and $\delta N^{(\rm s1)}$ is expanded up to the first order of $\delta\pi_1/\pi_1$.
It can be seen that $\delta N^{(\rm s1)}$ is strongly suppressed by $\Delta\varphi$ and $(\pi/\pi_1)^{1+\frac{6}{\eta_1}}$ compared to $\delta N^{(1)}$.
For this reason, we have ignored $\delta N^{(\rm s1)}$ in the main part of the paper.

The number of e-folds for the S2 region can then be calculated by imposing the same condition, i.e., $\Delta\varphi$ is small (quantitatively, $\Delta\varphi\ll\kappa g|\pi_1|$).
Since $\omega_{\rm s2}$ is considerably larger than $3/2$, the damping factor $e^{-3(n-n_2)/2}$ in Eq.~(\ref{appBGsolstep2}) is negligible. 
Under this approximation, Eq.~(\ref{appBGsolstep2}) can be solved inversely for $N^{(\rm s2)}\equiv N(\varphi_c,\pi_c;\varphi_2,\pi_2)=n_2-n_c$, and it yields 
\be
N^{(\rm s2)}\simeq\frac{1}{\omega_{\rm s2}}\sinh^{-1}\left(\frac{\Delta\varphi}{2|\pi_2|}\omega_{\rm s2}\right)
\simeq\frac{1}{\omega_{\rm s2}}\log\left(\frac{\Delta\varphi}{|\pi_2|}\omega_{\rm s2}\right)\,.
\label{appNs2}
\ee
The number of e-folds in the perturbed trajectory, ${\cal{N}}^{(\rm s2)}\equiv N(\varphi_c,\pi_c+\delta\pi_c;\varphi_2,\pi_2+\delta\pi_2)$, is also represented in the same form.
As a result, we obtain
\be
\delta N^{(\rm s2)}\simeq-\frac{1}{\omega_{\rm s2}}\log\left(1+\frac{\delta\pi_2}{\pi_2}\right)\,.
\label{appdeltaNs2}
\ee
It is important to note that this may not be just a correction term.
Certainly $1/\omega_{\rm s2}$ is of the order of $\Delta\varphi$, however when $\delta\pi_2$ is comparable to $-\pi_2$, $\delta N^{(\rm s2)}$ may diverge to infinity. 
This is a consequence of the existence of perturbed trajectories that barely reach the local maximum of the potential and take an enormous amount of time to pass through the step.
Therefore, the contribution to the total $\delta N$ from $\delta N^{(\rm s2)}$ cannot be ignored and must be taken into account.

\section{Hubble friction corrections during step stage}
\label{Hubblefriction}
\renewcommand{\theequation}{D.\arabic{equation}}
\setcounter{equation}{0}
In the main text we imposed the approximation condition (I\hspace{-1.2pt}I\hspace{-1.2pt}I) and neglected the Hubble friction during the step stage.
However, as we have seen in Appendix.~\ref{SRstages}, the ${\cal{O}}(\Delta\varphi)$ corrections are not necessarily negligible compared to the zeroth orders in the background solution, since the first and second terms in Eq.~\eqref{apppi22} largely cancel each other out for small $g$.
In this section we verify that they are negligible at the perturbation level and that Eq.~\eqref{deltapi2} is a good enough approximation.

From Eq.~\eqref{apppi22}, we can obtain
\ba
\pi_2+\delta\pi_2&=&\pi_2\biggl[1+\frac{2}{g^2}\frac{\delta\pi_1}{\pi_1}+\frac{1}{g^2}\left(\frac{\delta\pi_1}{\pi_1}\right)^2-\frac{3}{g^2}\left(1+\frac{\delta\pi_1}{\pi_1}\right)^2\delta N^{(\rm s1)}-\frac{3\delta\pi_1}{g^2\pi_1}\left(2+\frac{\delta\pi_1}{\pi_1}\right)N^{(\rm s1)}\notag\\
&&\hspace{1cm}+\frac{3\Delta\varphi}{g^2}\delta\pi_c-3\left(1+\frac{\delta\pi_2}{\pi_2}\right)^2(N^{(\rm s2)}+\delta N^{(\rm s2)})+3N^{(\rm s2)}\biggr]^{\frac{1}{2}}\,.
\label{appdeltapi2}
\ea
In the following, for simplicity, the second and higher orders of $\delta\pi_1/\pi_1$ are truncated. 
This is well justified for $g\ll1$.
Substituting the already calculated $\Delta\varphi$ zeroth order solutions~\eqref{appNs1}, \eqref{deltaNs1}, \eqref{appNs2}, \eqref{appdeltaNs2} and \eqref{deltapi2} into Eq.~\eqref{appdeltapi2}, we obtain
\ba
&&\pi_2+\delta\pi_2=\pi_2\biggl[1+\frac{2}{g^2}\frac{\delta\pi_1}{\pi_1}-\frac{3}{g^2}\delta N^{(\rm s1)}-\frac{6\delta\pi_1}{g^2\pi_1}N^{(\rm s1)}+\frac{3\Delta\varphi}{g^2 \pi_c}\frac{\delta\pi_1}{\pi_1}-\frac{6\delta\pi_1}{g^2\pi_1}N^{(\rm s2)}-3\left(1+\frac{\delta\pi_2}{\pi_2}\right)^2\delta N^{(\rm s2)}\biggr]^{\frac{1}{2}} \notag\\
&&\hspace{0.3cm}\simeq\pi_2\biggl[1+\frac{2}{g^2}\left(1-\frac{3}{2\omega_{\rm s1}}+\frac{3\Delta\varphi}{2\pi_c}-\frac{3}{\omega_{\rm s2}}\log\left(\frac{\sqrt{2}}{g}\right)\right)\frac{\delta\pi_1}{\pi_1}+\frac{3}{2\omega_{\rm s2}}\left(1+\frac{2}{g^2}\frac{\delta\pi_1}{\pi_1}\right)\log\left(1+\frac{2}{g^2}\frac{\delta\pi_1}{\pi_1}\right)\biggr]^{\frac{1}{2}}\,.
\ea
Therefore, if the step width is sufficiently small, i.e. $\omega_{\rm s1},\omega_{\rm s2}\gg1$ and $\omega_{\rm s2}\gg|3\log g|$, the $\Delta\varphi$ corrections are subdominant and negligible.
In this case, the Hubble friction can be ignored when calculating $\delta\pi_2$ with $g$ as an input parameter, resulting in Eq.~\eqref{deltapi2} being a good approximation.

\section{Derivation of PDF for each case}
\label{appeachPDF}
\renewcommand{\theequation}{E.\arabic{equation}}
\setcounter{equation}{0}
Using Eq.~(\ref{PDFofR}), we can numerically obtain the PDF of $\cal{R}$.
However, since it is not possible to solve Eq.~(\ref{R}) analytically, we cannot obtain the analytically explicit form of the PDF in terms of $\cal{R}$.
Instead of solving Eq.~(\ref{R}) exactly, it may be useful to derive the shape of the PDF in each case where one term in Eq.~(\ref{R}) makes a larger contribution than others, in order to understand the specific features of the full PDF in this model. 
For simplicity, in the following discussion we assume that $g\ll1$, otherwise the PDF in our model becomes close to the Gaussian case.

\begin{enumerate}
\item The linear perturbative regime\\
In this case, the PDF is obviously given by a Gaussian distribution with variance $\sigma_{\cal{R}}^2=(\beta-\kappa\gamma/3g-\gamma/\omega_{\rm s2}g^2)^2\sigma_{\delta\varphi}^2$, namely, 
\be
P[{\cal{R}}]=\frac{1}{\sqrt{2\pi\left(\beta-\frac{\kappa\gamma}{3g}-\frac{\gamma}{\omega_{\rm s2}g^2}\right)^2\sigma_{\delta\varphi}^2}}\exp\left(-\frac{{\cal{R}}^2}{2\left(\beta-\frac{\kappa\gamma}{3g}-\frac{\gamma}{\omega_{\rm s2}g^2}\right)^2\sigma_{\delta\varphi}^2}\right)\,.
\label{appGaussiandis}
\ee
\item The second term dominant regime\\
Based on the assumption $g\ll1$, we can ignore the third term in the square root in the second term of Eq.~(\ref{R}), i.e.
\be
{\cal{R}}\simeq\frac{\kappa g}{3}\left(1-\sqrt{1+\frac{2\gamma}{g^2}\delta\varphi}\right)\,.
\ee
The inverse of this equation gives us
\be
\delta\varphi\simeq\frac{9}{2}\frac{{\cal{R}}^2}{\gamma\kappa^2}-3\frac{g{\cal{R}}}{\gamma\kappa}\,,
\ee
and the PDF is 
\be
P[{\cal{R}}]=\frac{1}{\sqrt{2\pi \sigma_{\delta\varphi}^2}}\frac{3g}{|\gamma|\kappa}\left(1-\frac{3{\cal{R}}}{g\kappa}\right)\exp\left[-\frac{1}{2\sigma_{\delta\varphi}^2}\frac{9g^2{\cal{R}}^2}{\gamma^2\kappa^2}\left(1-\frac{3}{2}\frac{{\cal{R}}}{g\kappa}\right)^2\right]\,.
\label{appPDFcutoff}
\ee
We note that in this case there is a cutoff at ${\cal{R}}=\kappa g/3$.
It can be seen that the PDF decreases rapidly towards 0 at ${\cal{R}}=\kappa g/3$ due to the factor $1-3{\cal{R}}/g\kappa$.
Since the cutoff depends not only on $\kappa$ but also on $g$, the smaller $g$ we choose, the smaller the cutoff tends to be.
\item The third term dominant regime\\
Around $\delta\varphi\simeq-2\gamma/g^2$, the main contribution to $\cal{R}$ is the third term in Eq.~(\ref{R}), namely,
\be
{\cal{R}}\simeq-\frac{1}{2\omega_{\rm s2}}\log\left(1+\frac{2\gamma}{g^2}\delta\varphi\right)\,.
\ee
Thus $\delta\varphi$ can be given by 
\be
\delta\varphi\simeq\frac{g^2}{2\gamma}\Bigl(\exp\left(-2\omega_{\rm s2}{\cal{R}}\right)-1\Bigr)\,,
\ee
and then the PDF is 
\be
P[{\cal{R}}]=\frac{1}{\sqrt{2\pi\sigma_{\delta\varphi}^2}}\frac{g^2 \omega_{\rm s2}}{|\gamma|}\exp\left(-2\omega_{\rm s2}{\cal{R}}\right)\exp\left[-\frac{1}{2\sigma_{\delta\varphi}^2}\frac{g^2}{4\gamma^2}\Bigl(\exp\left(-2\omega_{\rm s2}{\cal{R}}\right)-1\Bigr)^2\right]\,.
\label{appPDFoftail}
\ee
For a sufficiently large ${\cal{R}}$, this PDF is almost proportional to $\exp(-2\omega_{\rm s2}{\cal{R}})$, i.e. the tail appears.
Recalling that $\omega_{\rm s2}\propto1/\Delta\varphi$, we can see that the effect of this tail is suppressed by $\Delta\varphi$.
If we take the case of zero width limit, $\Delta\varphi=0$, then the tail vanishes and the cutoff described in Refs.~\cite{Cai:2021zsp,Cai:2022erk} can be reproduced.
However, the PDF is nonzero even for ${\cal{R}}$ larger than the cutoff value, ${\cal{R}}_{\rm cutoff}=\kappa g/3$, unless $\Delta\varphi$ is exactly zero.
In this sense, we can conclude that the cutoff is a side effect of ignoring the finite step width.
\end{enumerate}

\bibliographystyle{mybibstyle}
\bibliography{bib}

\end{document}